\newcommand{\be}{\begin{eqnarray}}
\newcommand{\ee}{\end{eqnarray}}

\renewcommand{\thefootnote}{\fnsymbol{footnote}}
\newcommand{\vecb}[1]{{\bf #1}}
\newcommand{\vecg}[1]{\mbox{\boldmath $#1$}}
\newcommand{\lsim}{\lesssim}

\documentclass[12pt]{article}

\usepackage{amsmath,amssymb}
\usepackage{graphicx}
\usepackage{cite}
\usepackage{bm}

\def\theequation{\arabic{section}.\arabic{equation}}
\begin{document}

\vskip 15mm

\begin{center}

{\Large Witten index in supersymmetric $3d$ theories revisited}

\vskip 4ex

A.V. \textsc{Smilga}\,$^{1}$,

\vskip 3ex

$^{1}\,$\textit{SUBATECH, Universit\'e de
Nantes,  4 rue Alfred Kastler, BP 20722, Nantes  44307, France
\footnote{On leave of absence from ITEP, Moscow, Russia.}}
\\
\texttt{smilga@subatech.in2p3.fr}
\end{center}

\vskip 5ex

\begin{abstract}
\noindent  We have performed a direct calculation of Witten index $I$
in ${\cal N} = 1,2,3$ supersymmetric  Yang-Mills Chern-Simons (SYMCS) $3d$ theories. 
We do it in the framework of Born-Oppenheimer (BO) approach by putting the system 
into a small spatial box  and studying the effective Hamiltonian depending on the zero field harmonics.
At the tree level, our results coincide with the results of Ref.\cite{Wit99}, but there is
a difference in the way the loop effects are implemented. In Witten's approach, one has only take
into account the fermion loops, which bring about a negative shift of the (chosen positive at the tree level) 
Chern-Simons coupling $k$. As a result, Witten index vanishes and supersymmetry is broken at small $k$. 
In the effective BO Hamiltonian framework, fermion, gluon and ghost loops contribute on an equal
footing. Fermion loop contribution to the effective Hamiltonian can be evaluated exactly, and their effect 
amounts to the negative shift  $k \to k - c_V/2$ for ${\cal N} =1$ and $k \to k - c_V$ for ${\cal N} = 2,3$
in the tree-level formulae for the index. In our approach, with rather natural assumptions on the structure of bosonic corrections, 
the shift $k \to k + c_V$ brought about by the gluon loops also affects the index.
 Since the total shift of $k$ is positive or zero, Witten index appears to be nonzero at  nonzero $k$, 
and supersymmetry is not broken.

We discuss possible  reasons for such disagreement 

\end{abstract}

\renewcommand{\thefootnote}{\arabic{footnote}}
\setcounter{footnote}0
\setcounter{page}{1}

\section{Introduction}
 It is known since \cite{Malda} that ${\cal N} = 4$ supersymmetric Yang-Mills theory in 4 
dimensions is dual to supersymmetric string theory ($10d$ supergravity in the leading strong
 coupling approximation) on $AdS_5 \times S^5$ background. In other words, many nontrivial
 results for ${\cal N} = 4$ SYM theory for large $N_c$ and  large 't Hooft coupling can be 
obtained by string theory methods. Recently, a new interesting duality has been established. 
It relates certain $3d$ supersymmetric gauge theories, involving the Chern-Simons terms and a
 particular set of matter fields and  enjoying ${\cal N} = 8$ or ${\cal N} = 6$ supersymmetry,
 to string theories on $AdS_4 \times S^7$ or $AdS_4 \times \mathbb C \mathbb P^3$ backgrounds, respectively 
\cite{3duality}. This means that, by duality, one can derive many nontrivial results for these $3d$ theories.

 The theories in question are not so simple, and we do not understand their dynamics as well as we 
do it for $4d$ theories. In our opinion, it makes sense to study it in as much details as possible 
by {\it purely} field theory methods in order to be able to confront the results thus obtained 
with the results following from string-gauge duality. A wish to develop tools that would eventually 
allow us  to perform such a comparison and to test the duality conjecture once again was the main
 motivation behind the present study.

 As was mentioned, the $3d$ theories, for which duality was established, are complicated. Thus, we 
have decided to study first the simplest ${\cal N} =1$ SYMCS theory and, in particular, its vacuum 
dynamics. This question was addressed previously in Ref.\cite{Wit99}. Witten calculated the index 
(the difference of the numbers of bosonic and fermionic vacuum states) for this theory. His result 
for the theory with $SU(N)$ gauge group at the level $k = \kappa/(4\pi) $ is
  \be
  \label{IndpoWittenu}
  I(k,N) \ =\  \frac 1{(N-1)!} \prod_{j = -\frac N2 + 1}^{\frac N2 -1} (k-j) \ .
 \ee
This is zero at $|k| < N/2$. For $|k| \geq N/2$, it can be presented as 
  \be
I(k,N) \ =\  
(-1)^{N-1} \left( \begin{array}{c} |k|+N/2 -1 \\ N-1 \end{array} \right)\ .
 \ee
 The way this result was derived was not direct, however.  That is why we have tried to evaluate the index anew
 using more direct and clear physical reasoning. We use the same method as Witten successfully applied in \cite{Wit82} 
for $4d$ supersymmetric gauge theories: put the system in a small spatial box and impose periodic
 boundary conditions on all fields. If the size of the box is made small enough, most of the variables
 in the field Hamiltonian become {\it fast} with large characteristic excitation energies. One can 
integrate them over and study the dynamics of the effective BO Hamiltonian that depends only on few 
{\it slow} variables (zero Fourier modes of gauge fields belonging to the Cartan subalgebra and their superpartners).

 However, it turns out that carrying out this program for $3d$ SYMCS theories is a more difficult task 
than for $4d$ gauge theories. It might even seem that it {\it fails} in the $3d$ case because it is 
not sufficient
 to restrict oneself here with the tree-level effective Hamiltonian. Loop corrections are important and they 
change essentially the value of the index. At the one-loop level, these corrections can be determined, 
however, and one can { conjecture} that higher-loop effects do not further change
 the result. This conjecture is not quite proven by now, but, following Witten, we find it plausible 
(the arguments in its favor will be discussed later) and adopt it.

As we will see, the index of the effective finite volume BO Hamiltonian depends on the $r$-th Chern class of 
a certain  Abelian gauge field on the moduli space of flat connections, with $r$ being the rank of the group.
 In the case of $SU(2)$, it is just the magnetic field flux on the dual torus. 
 There are one-loop contributions to this (generalized) flux, both due to fermion loops and due to gluon loops.
These corrections are associated with the renormalization of the Chern-Simons coefficient $k$ in the infinite volume
theory. Thus, the index can be evaluated in two steps. 
 \begin{itemize} 
 \item At the first step, one evaluates the index for the tree-level effective BO Hamiltonian.
We have performed it by another method than Witten and confirmed his result,
\be
\label{nashindextree}
I^{\rm tree}(k,N) \ =\ \left( \begin{array}{c} k+N -1 \\ N-1 \end{array} \right) \ 
 \ee
(This is for the $SU(N)$ gauge group and positive $k$).  
 \item At the second step, one takes into account loop effects, which boil down (we will argue that later) 
to 1-loop renormalization of  $k$ due to both fermions and bosons,
 \be
\label{shiftk}
 k_{\rm ren} \ =\ k - \frac {c_V}2 ({\rm fermions})  + c_V ({\rm bosons}) \ = \ k+ \frac {c_V}2 \ ,  
  \ee
\end{itemize}
where $c_V$ is the adjoint Kasimir eigenvalue. [For $SU(N)$, $c_V = N$. For $Sp(2r)$, $c_V = r+1$. Another name for $c_V$ is the
dual Coxeter number $h$.]  
Substituting (\ref{shiftk}) in (\ref{nashindextree}), we obtain [$SU(N)$, positive $k$]
  \be
  \label{nashindex}
  I (k,N)  \ = \    \left( \begin{array}{c} k+ 3N/2 -1 \\ N-1 \end{array} \right) \ ,
  \ee
Note that one would obtain Witten's result (\ref{IndpoWittenu}) by doing the same, but leaving only the 
contribution of the fermionic loops in (\ref{shiftk}). The fact that gluon loops contribute to the shift of $k$ is firmly
established \cite{Rao,Kao}. It is less clear, however, whether such boson-induced shift of $k$  is directly translated 
into the shift of index. In Witten's approach, it does not. 
 In our finite volume approach, a direct and {\it quite} honest evaluation of the gluon contribution to the effective BO Hamiltonian
 is, technically, a more complicated problem than for the fermion contribution (the latter can be evaluated exactly), which is 
still to be solved. But 
under very natural assumptions, the boson contribution has the same structure as the fermion one. The result (\ref{nashindex}) is obtained
under {\it this} assumption.

The difference between (\ref{nashindex}) and (\ref{IndpoWittenu}) is essential. 
The product (\ref{IndpoWittenu}) vanishes
at $k < N/2$, which suggests spontaneous breaking of supersymmetry. 
But the expression (\ref{nashindex}) does not display
such feature meaning that supersymmetry is not broken. 
Neither is it broken in ${\cal N} = 2,3$ theories, where fermion loop
and gluon loop effects in the renormalization of $k$  cancel out (scalar loops contribute to 
renormalization of $g^2$, but not
to renormalization of $\kappa$), and the index is given by the 
tree level expression (\ref{nashindextree}).
We will discuss this controversy in more details in the last section.

 In the next section, we fix  notations and calculate the index at the tree level. The index of the 
original theory is evaluated as the index of the effective SQM Hamiltonian, where one should impose the 
additional constraint of {\it Weyl invariance} of wave functions (this is a corollary of gauge invariance 
of wave functions in the full theory). Before this restriction is imposed, one finds $Nk^{N-1}$ vacuum 
states for the $SU(N)$ gauge group. The wave functions of all these states can be explicitly  determined: 
they represent generalized theta functions. Not all these functions are invariant under Weyl transformations, 
however, the total number of Weyl-invariant functions being given by the expression (\ref{nashindextree}).
 
 We also calculate the  index for the symplectic gauge groups $Sp(2r)$ and for $G_2$. For symplectic groups, the calculation is 
 even more transparent  
 than for unitary groups. The (tree level) result is
 \be
 \label{indSp}
 I[{\rm Sp}(2r)] \ =\ \left( \begin{array}{c} k+r \\ r \end{array} \right)\ .
  \ee
The result for $G_2$ is
\be
 \label{indG2}
 I^{\rm tree}_{G_2}( k) \ =\ \left\{ 
                 \begin{array}{c}  \frac {(|k|+2)^2}4\  \ \ \ \ \ \ \ \ {\rm for\ even} \  k \\
                                   \frac {(|k|+1)(|k|+3)}4 \ \ \ \ \ \ \ \ {\rm for\ odd} \  k 
 \end{array}
               \right\} \ .
\ee

 In Sect. 3 we discuss one-loop corrections. We show  that they amount to
shifting $k$, as dictated by (\ref{shiftk}). 
 We discuss also the ${\cal N} = 2,3$ SYMCS $3d$ theories including  extra adjoint Majorana fermions and 
 extra adjoint real scalars, and show that the index there is just given  by Eqs.(\ref{nashindextree}), (\ref{indSp})
with unshifted $k$.  

Sect. 4 is devoted to discussions. We spell out again the reasoning leading to the 
result (\ref{nashindex}) and confront it with Witten's reasoning. In addition, we address the unclear by now question of what might be
 wrong with the string-inspired arguments of Ref.\cite{Vafa}, which favor the result (\ref{IndpoWittenu}) 
rather than (\ref{nashindex}).

\section{Tree level}
\setcounter{equation}0

 The action of ${\cal N} =1$ SYMCS theory is 
\footnote{See e.g. \cite{Dunne} for a nice review.}
 \be
 \label{LN1}
  {\cal L} \ =\ \frac 1{g^2} {\rm Tr} \int d^3x \left \{ - \frac 12 F_{\mu\nu}^2 +
  i\bar \lambda /\!\!\!\!D \lambda \right \} +
  \kappa {\rm Tr}\int d^3x  \left\{ \epsilon^{\mu\nu\rho}
  \left( A_\mu \partial_\nu A_\rho - \frac {2i}3 A_\mu A_\nu A_\rho \right ) - \bar \lambda \lambda \right \}
   \ee
with the conventions $\epsilon^{012} = 1, \ D_\mu {\cal O}  = \partial_\mu {\cal O}  - i[A_\mu, {\cal O}] $;  
   $\lambda_\alpha$ is a 2-component Majorana $3d$ spinor belonging to the adjoint representation of the gauge group.
   We choose
   \be
   \label{gamdef}
   \gamma^0 \ =\ \sigma^2,\ \ \ \gamma^1 = i\sigma^1,\ \ \ \gamma^2 = i\sigma^3 \ .
    \ee
 This is  a $3d$ theory and the coupling constant $g^2$ carries the dimension of mass. The physical 
boson and fermion degrees of freedom in this theory are massive,
  \be
  \label{mass}
  m = \kappa g^2\ .
   \ee
 In three dimensions, the nonzero mass brings about parity breaking. 

  The parameter $\kappa$ is dimensionless. It cannot be an arbitrary number, however. The functional integral should 
be invariant with respect to {\it large} gauge transformations that change the Chern-Simons number of the gauge field configuration,
 \be
\label{NCS}
N_{CS} \ =\  \frac 1{8\pi^2} \epsilon^{\mu\nu\rho} {\rm Tr}  \, \int d^3x\,  
  \left( A_\mu \partial_\nu A_\rho - \frac {2i}3 A_\mu A_\nu A_\rho \right )
 \ee
by an integer. The requirement for $e^{iS}$ to be invariant under such transformation leads to the quantization condition
  \be
    \label{quantkap}
    \kappa =  \frac k {4\pi}  \ .
    \ee
with integer $k$. 
Two reservations are in order, however. First, we consistently assume in this paper that the field theory (\ref{LN1}) 
is regularized in the infrared by putting it on 
a spatial torus with 
{\it periodic} boundary conditions. 
If the so called {\it twisted} boundary conditions were imposed \cite{tHooft}, Chern-Simons 
number could change by an integer multiple of $1/N$,
in which case $k$ would be 
 quantized to be an integer multiple of $N$ \cite{Wit99}. Second, we have not taken into account loop effects  yet. 
We shall learn in Sect. 3 that the loops may in some cases modify the quantization condition such that $k$  must be half-integer.

    The parameter $k$ is called the {\it level} of the theory.

\subsection{Effective Hamiltonian}

We put the system in a spatial box of size $L$ and impose periodic boundary conditions on the fields. The Witten
 index does not depend on the size of the box and we are allowed to consider the limit
 \be
 \label{smallbox}
  mL \ll 1 \ \ \ \ \ \ {\rm and\ hence} \ \ \ \ \ \ g^2L \ll 1 \ .
  \ee
  [The second inequality follows from the first one, from the definition (\ref{mass}) and from the quantization 
condition (\ref{quantkap})].
We expand the dynamic field variables in the Fourier series.
 \be
 \label{Fourier}
 A_j(\vecb{x}) &=& \sum_{\vecb{n}} A_j^{(\vecb{n})} e^{2\pi i \vecb{x} \vecb{n}/L} \ \ \ \ \ \ \ \ \ \ \ (j=1,2) , \nonumber \\
\lambda_\alpha(\vecb{x}) &=& \sum_{\vecb{n}} \lambda_\alpha^{(\vecb{n})} e^{2\pi i \vecb{x} \vecb{n}/L}
 \ee
 with integer $\vecb{n}$. When the condition (\ref{smallbox}) is satisfied,
 the zero Fourier components $A_j^{(\vecb{0})}$ and $\lambda_\alpha^{(\vecb{0})}$
belonging to the Cartan subalgebra of the full Lee algebra play a special role: the characteristic excitation 
energies associated with these degrees of freedom are of order $E^{(0)} \sim g^2$, which is much less than the 
characteristic excitation energy $E^{\rm higher\ modes} \sim 1/L$ associated with higher Fourier harmonics and 
much less than the characteristic energy associated with non-Abelian components of the vector potential
$E^{\rm non-Ab} \sim (g/L)^{2/3}$. We can thus integrate over the fast variables $A_j^{(\vecb{n} \neq \vecb{0})}$, etc. 
and build up the effective BO Hamiltonian (and the corresponding Lagrangian) depending only on the slow variables 
$A_j^{(\vecb{0})\ {\rm Cartan}}$ and $\lambda_\alpha^{(\vecb{0})\ {\rm Cartan}}$. The situation is exactly the same
 as
for $4d$ theories \cite{Wit82}.

In the tree approximation, the effective Lagrangian is obtained by a simple truncation of
all fast modes in (\ref{LN1}). Proceeding in a similar way for $4d$ theories, we would obtain the 
Lagrangian/Hamiltonian describing free motion on $T\times T \times T$, with
$T$ representing the maximal torus of the group \cite{Wit82}. In the $3d$ case, the situation is more complicated.

Consider first the simplest $SU(2)$ case. There are two slow bosonic variables
 \be
 \label{defslow}
C_j \equiv A_j^{(\vecb{0})\, 3}
 \ee
 and their superpartners $\psi_\alpha \equiv \lambda_\alpha^{(0)\, 3}$.
The truncated Lagrangian is
 \be
 \label{Leffsimplest}
 {\cal L} \ =\ \frac {L^2}{2g^2} \dot{C}_j^2 - \frac {\kappa L^2}2 \epsilon_{jk} C_j \dot{C}_k + 
\frac {iL^2}{2g^2}\psi_\alpha \dot{\psi}_\alpha + \frac {i\kappa L^2}2 \epsilon_{\alpha \beta} \psi_\alpha \psi_\beta
  \ee
  To  find the corresponding Hamiltonian, it is convenient to introduce
  $\psi_{\pm} = \psi_1 \pm i \psi_2$. Then the fermion part of the Lagrangian is represented as
   \be
   {\cal L}_f \ =\ \frac {iL^2}{2g^2}  \psi_+ \dot{\psi}_-    + \frac {\kappa L^2}2 \psi_- \psi_+ \ .
    \ee
    We see that the only fermion dynamic variable is $\psi_- \equiv \psi$. Note that it is transformed as
    \be
    \label{psirotat}
    \psi \to e^{i\theta/2} \psi
      \ee
   under spatial plane rotations. The canonical momentum is $\pi_\psi = iL^2\psi_+/(2g^2)$.
    After quantization, it goes over  to $-i \partial/ \partial \psi \equiv -i\bar \psi$. Ordering the product 
$\bar\psi \psi$ in a proper (Weyl) way and introducing also bosonic canonical momenta $P_j$, we derive the quantum Hamiltonian
    \be
    \label{Heffsimplest}
    H \ =\ \frac {g^2}{2L^2} \left(P_j - \frac {\kappa L^2}2 \epsilon_{jk} C_k \right)^2 + \frac {\kappa g^2}2 
(\psi \bar\psi - \bar\psi \psi)\ .
    \ee
    It describes the motion in the presence of a uniform magnetic field $B = \kappa L^2$ on the dual 2-dimensional 
torus $C_{j=1,2} \in (0, 4\pi/L)$.
   The motion is finite because all the points $C_j + 4\pi n_j/L$ with integer
$n_j$ are gauge-equivalent.

   The motion of electron in a uniform magnetic field is the first and the simplest supersymmetric quantum problem ever 
considered \cite{Landau}. The bosonic and fermionic sectors of the Hamiltonian (\ref{Heffsimplest}) correspond in the
 usual approach to spin-up and spin-down  electrons. The index of this Hamiltonian $I = {\rm Tr} \{ (-1)^F e^{-\beta H} \}$ 
can be calculated as a functional integral, which is reduced for small $\beta$
  (semiclassical limit) to an ordinary phase space integral \cite{Cecotti}
  \footnote{In the Hamiltonian (\ref{Heffsimplest}), the magnetic field is constant, but, in view of future applications, 
we write the index for a generalized Hamiltonian describing the motion in a non-uniform magnetic field \cite{Novikov}.}.
   \be
   \label{indCecotti}
   I \ =\ \int \prod_{j=1,2} \frac {dP_j dC_j}{2\pi} \, d\bar\psi d\psi\
   e^{-\beta H} \ = \frac 1{2\pi} \int  {\cal B}(\vecb{C}) \  d\vecb{C}   \ .
   \ee
   When the motion extends over the whole plane, the index is infinite, indicating the infinite ground state degeneracy. 
When the motion is finite, the number of vacuum states is finite, 
   being proportional to the total magnetic flux. In our case, ${\cal B} =
   \kappa L^2$ and
    \be
    \label{ind2avant}
I = \frac {\kappa L^2}{2\pi} \left( \frac {4\pi}L \right)^2 = 8\pi\kappa = 2k \ .
 \ee

\begin{figure}[t]
\begin{center}
\includegraphics[width=3in]{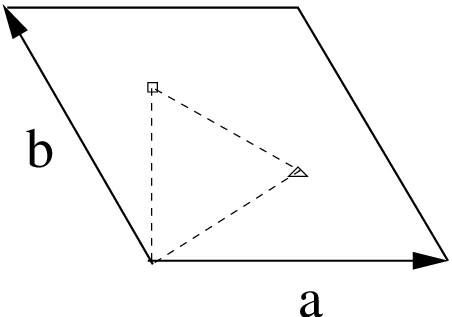}
\end{center}
\caption{Maximal torus and Weyl alcove for $SU(3)$. $\vecb{a}$ and $\vecb{b}$ are simple coroots. 
The points $\Box$ and $\triangle$ are fundamental coweights.}
\label{romb}
\end{figure}

  It is not difficult to  generalize this analysis to other gauge groups.  In general,  we have $2r$ slow bosonic 
variables $C_{ja}$ and their superpartners $\psi_a$. The index $a= 1,\ldots,r$ labels the generators of the Cartan 
subalgebra with the usual convention Tr$\{t^a t^b\} = \delta^{ab}/2$. The effective Hamiltonian belongs to the class 
of supersymmetric Hamiltonians
     \be
    \label{HeffN}
    H \ =\ \frac {g^2}{L^2} \left[ \frac {(P_{ja} + {\cal A}_{ja})^2}2 + \frac 12 {\cal B}_{ab}
 (\psi_a \bar\psi_b - \bar\psi_b \psi_a) \right] \ ,
    \ee
    where ${\cal B}_{ab} = \epsilon_{jk} \partial_{aj} {\cal A}_{bk} $, describing a generalized 
multidimensional  
Landau-Dubrovin-Krichever-Novikov problem. For the tree-level Hamiltonian that corresponds to the truncated
 Lagrangian of Eq.(\ref{LN1}),
     \be
     \label{postojan}
     {\cal A}_{aj} &=&   -\frac {\kappa L^2}2 \epsilon_{jk} C_{ak} \ , \nonumber \\
     {\cal B}_{ab} &=& \kappa L^2 \delta_{ab} \ .
    \ee
    By the same token as in the $SU(2)$ case, the motion is finite and extends for each $\vecb{C}$   over 
a parallelepiped  formed by simple coroots of the group (alias, the maximal torus $T$ of the group).
 For $SU(3)$, this is a rhombus  represented in Fig.\ref{romb} 
(do not pay attention for a while to the dashed lines bounding the Weyl alcove,
 neither to special fundamental coweight points  marked by the box and triangle).

    The index of the effective Hamiltonian is evaluated semiclassically as a {\it generalized} magnetic flux
(this is nothing but that the r-{\rm th} Chern class of the $U(1)$ bundle over $T \times T$ with the connection ${\cal A}_{ja}$),
   \be
    \label{indCecotti3}
    I =  \frac 1{(2\pi)^r} \int_{T \times T} \prod_{ja} dC_{ja} \ {\rm det} \| {\cal B}_{ab} \| \ .
     \ee
    In the case of $SU(N)$,
    \be
    \label{ISUN}
    I^{SU(N)} \ =\ N k^{N-1}\ .
     \ee

\subsection{Counting Weyl invariant vacuum functions}

 The index of the effective Hamiltonian (\ref{HeffN}), (\ref{postojan}) is given by the expression (\ref{ISUN}). 
But the index of the original theory {\it is} not. There are two reasons by which the result (\ref{ISUN}) is 
modified. The first reason (loop effects) was already mentioned. We will deal with loops in the next section. 
The second reason is that the Schr\"odinger equation with the effective Hamiltonian (\ref{HeffN}) should in fact be 
supplemented by the condition 
of Weyl invariance, which is a corollary of the gauge invariance of the original theory \cite{Wit82}. For example, 
for $SU(2)$, wave functions should be invariant under the reflection $C_j \to -C_j,\ \ \ \psi \to
 -\psi$. Not all eigenfunctions of (\ref{HeffN}) satisfy this requirement. As a result, the value of the index is 
less than  "pre-Weyl" index (\ref{ISUN}).

 To find it, we simply write down explicit expressions for all vacuum wave functions and pick up Weyl-invariant ones. 
To begin with, consider the simplest $SU(2)$ case and let first $k$ be positive. The ground states of the effective
 Hamiltonian have then zero fermion charge such that the second term in the Hamiltonian (\ref{Heffsimplest}) brings
 about a negative contribution to the energy.

 Let us introduce $x = C_1 L/(4\pi) \in (0,1)$ and $y = C_2 L/(4\pi) \in (0,1)$. All eigenfunctions of the Hamiltonian
 satisfy the following boundary conditions
\be
\label{bc2}
\Psi(x+1, y) &=& e^{-2\pi i ky} \Psi(x,y) \ , \nonumber \\
\Psi(x, y+1) &=& e^{2\pi i kx} \Psi(x,y) \ .
 \ee

Their origin can be traced back to the fact that the shifts $ x \to x+1$ and 
$y \to y+1$ represent {\it contractible} (this is the non-Abelian specifics) gauge transformations.
 In most gauge theories, wave functions are invariant under such transformations. 
But the YMCS (or Maxwell + CS) theory is special in this respect \cite{Deser}.
Indeed, the Gauss law constraint in the YMCS theory has the form
 $$ G^a = \frac {\delta {\cal L}}{\delta A_0^a} = D_j \Pi_j^a + \frac \kappa 2 \epsilon_{jk}
 \partial_j A^a_k \ , $$
 where $\Pi_j^a = F_{0j}^a/g^2 + (\kappa/2) \epsilon_{jk}
  A^a_k $ are the canonical momenta. The second term gives rise to the phase factor associated 
with an infinitesimal gauge transformation $\delta A^a_j (\vecg{\xi})  = D_j \alpha^a (\vecg{\xi})$ 
(we denote here the usual spatial coordinates by $\vecg{\xi}$ rather than $\vecg{x}$ not to confuse 
them with rescaled vector potentials), 
 \be
\label{fazashift}   
 \Psi[A^a_j + D_j \alpha^a] \ =  \ 
\exp\left\{-\frac {i\kappa}2 \int d\vecg{\xi} \, \epsilon_{kl} \, \partial_k 
\alpha^a A^a_l \right \} \Psi[A^a_j] \ . 
  \ee
This property holds also for the finite contractible gauge transformations 
$\alpha^a =  (4\pi \xi_{1,2}/L) \delta^{a3}$ implementing the shifts $C_{1,2} \to C_{1,2} + 4\pi/L$. 
The phase factors thus obtained coincide with those quoted in  Eq. (2.19); they 
 are nothing but the holonomies $\exp \left\{i \int_0^{4\pi/L} {\cal A}_1 dC_1 \right\}$ and
 $\exp \left\{i\int_0^{4\pi/L} {\cal A}_2 dC_2\right\}$. Thereby, the eigenfunctions of $H$ are elliptic
 functions --- a variety of theta functions. The $2k$ ground states can be chosen in the form
   \be
   \label{Psim}
   \Psi_m \sim \ \sum_n \exp \left \{ -2\pi k \left(n+y + \frac m{2k}\right)^2- 2\pi i k x y -4\pi i k x 
\left(n + \frac m{2k} \right)
   \right\} \ ,
    \ee
    where the sum runs over all integer $n$, and $m = 0,\ldots,2k-1$.
    Not all of these states are invariant, however, under Weyl reflection
    $\{x \to -x, \ y \to -y\}$. There are only  $k+1$ Weyl invariant combinations:
 $$ \Psi_0,\ \Psi_k, \  {\rm and} \ \ \Psi_m + \Psi_{2k-m} \ \ (m = 1,\ldots,k-1) \ .$$
 In other words, the (tree-level) index is
   \be
   \label{indtree2}
   I^{\rm tree} [SU(2)] \ =\ k+1 \ .
    \ee
This explicit analysis was done for the constant magnetic field. However, the symmetry properties of
the wave functions are robust with respect to deformations. We thus can be sure that the number of Weyl-invariant
wave functions is equal to $k+1$ also for the Hamiltonian with nonuniform magnetic field of a given flux $2k$. 

\vspace{2mm}

\centerline{\it Fast Hamiltonian and its ground state.}
\vspace{2mm}

What happens at negative nonzero $k$ ? The ground states of the effective Hamiltonian (\ref{HeffN})
 are in this case not bosonic, but fermionic, involving $\psi$ as a factor. This factor is odd under Weyl 
reflection. At first sight, to provide for Weyl-evenness of the wave function, this should be compensated by
 picking up Weyl-odd combinations of the functions (\ref{Psim}). There are $|k| - 1$ such combinations 
which would lead to the conclusion that
 the index is equal to $k +1$ also for negative $k$ (giving $|k| - 1$ fermionic states). This is obviously
 wrong, however, the number of vacuum states cannot depend on the sign of $k$. To resolve this paradox, one
 should go into some details of the BO procedure.

When $k$ is positive, the wave functions (\ref{Psim}) are the ground states of the effective Hamiltonian 
(\ref{HeffN}). They depend  on the slow variables $C_{1,2}$ and the factor $\psi$ is absent in this case --- 
the states are bosonic.  The corresponding ground states of the {\it full} Hamiltonian are obtained when $\Psi_m$ 
are multiplied by the ground states of the  fast Hamiltonian depending on all Fourier modes (\ref{Fourier}) 
of the charged (with respect to $A^{(0)\,3}$) fields $A^{a=1,2}_i(\vecb{x})$ and $\lambda^{a=1,2}(\vecb{x}) 
\equiv
\lambda^{a=1,2}_1(\vecb{x}) - i\lambda^{a=1,2}_2(\vecb{x})$. The fault in the argument above (leading to the
 paradoxical result $I(k<0) = k+1$) does not depend, however, on the presence
of higher Fourier modes, and it is sufficient to analyze the dimensionally reduced theory where the fields 
do not depend on $\vecb{x}$. Let us assume that
 \be
 \label{Cj1}
C_j = C \delta_{j1}
 \ee
 and $C \gg m = \kappa g^2$. Then the fast Hamiltonian (in the quadratic with respect to fast variables 
approximation) acquires the form
 \be
 \label{Hfast}
H^{\rm fast} \ =\ \frac {g^2}2 \left( P^a_j - \frac \kappa 2 \epsilon_{jk} A^a_j \right)^2
+ \frac {C^2}{2g^2} \left( A^a_2 \right)^2 + \nonumber \\
\frac {iC}{4g^2} \epsilon^{ab} \left( 4g^4 \bar \lambda^a \bar \lambda^b + \lambda^a \lambda^b \right) + 
\frac m2 \left( \lambda^a \bar \lambda^a - \bar\lambda^a \lambda^a \right) \ ,
 \ee
where we have set for simplicity $L = 1$, and the index $a$ takes two "transverse" values, $a = 1,2$.
 Let us look first at the bosonic part. For each $a$, it describes the motion of a scalar particle in the
 magnetic field ${ B} = \kappa$ with an additional oscillatoric potential
 $\propto ( A^a_2)^2 $. The spectrum of a generic such Hamiltonian,
 \be
 \label{Bosc}
 H \ =\ \frac 1{2M} \left(p_x - \frac {By}2 \right)^2 + \frac 1{2M} \left(p_y + \frac {Bx}2 \right)^2 + 
\frac 12 \left( \omega_1^2 x^2 + \omega_2^2 y^2 \right)\ ,
  \ee
 is well known \cite{CMP},
 \be
 \label{specbosfast}
 E_{nl} \ =\ \left(n + \frac 12 \right) \Omega_1 + \left(l + \frac 12 \right) \Omega_2
 \ee
 with
 \be
 \label{Om12}
 \Omega_{1,2} \ =\ \frac 1{2M} \left[ \sqrt{B^2 + M(\omega_1 + \omega_2)^2} \pm
  \sqrt{B^2 + M(\omega_1 - \omega_2)^2} \right] \ .
   \ee
   In the case under consideration, 
   \be
   \label{nashiOm12}
   \Omega_1 = \sqrt{C^2 + m^2} \ , \ \ \ \ \ \ \ \Omega_2 = 0 \ .
    \ee
    The presence of two zero modes (as was mentioned above, $H^{\rm fast}_{\rm bos}$ represents the 
sum of two identical Hamiltonians  for $a=1,2$) is very natural. They are none other than the 
{\it gauge} modes corresponding to the action of the Gauss constraints operators $G^a$ on the vacuum 
and all other physical wave functions. If resolving explicitly the Gauss law constraints and expressing 
everything in terms of physical gauge-invariant variables, the zero modes associated with gauge 
rotations disappear. It is convenient, however, to leave
    the constraints unresolved. The bosonic vacuum wave function has then the form
     \be
     \label{Psivacbos}
     \Psi^{\rm fast}_{\rm bos} \ =\ \exp\left\{ - \frac {i\kappa}2 A_1^a A_2^a - 
\frac {\sqrt{C^2 + m^2}}{2g^2} \left( A_2^a \right)^2 \right\}
      \ee
It is annihilated by the operator $G^3$. The vanishing of $G^{1,2} \Psi$ is not explicit, 
but that is because the operators $G^{1,2}$ mix $A^a_j$ and $C_j$ ,
while (\ref{Psivacbos}) was written in the assumption that the slow bosonic variables have only
 the third color component. The corresponding eigenfunctions of the full bosonic Hamiltonian depend 
only on gauge-invariant combinations, like  $R_{jk} = \sum_{a=1}^3 A^a_j A^a_k$ and are annihilated by all three constraint operators.

The wave function (\ref{Psivacbos}) is multiplied by the ground state of the fermionic part of the 
Hamiltonian (\ref{Hfast}),
   \be
     \label{Psivacferm}
     \Psi^{\rm fast}_{\rm ferm} \ =\
     4ig^2  + \frac{ \sqrt{C^2 + m^2} - m }C \, \epsilon^{ab} \lambda^a \lambda^b \ .
      \ee
     The total energy is zero as it should: the contribution $\sqrt{C^2 + m^2}$ of the bosonic 
part cancels the fermionic contribution $-\sqrt{C^2 + m^2}$.
    The Hamiltonian (\ref{Hfast}) and the wave functions (\ref{Psivacbos}, \ref{Psivacferm})
were written in the assumption (\ref{Cj1}). It is equally easy to write them for arbitrary $C_j$.
 We will only need the expression for the fermion wave function:
     \be
     \label{PsivacfermanyC}
     \Psi^{\rm fast}_{\vecb{C}} \ =\
     4ig^2  + \ \frac {(C_1 - iC_2) \left( \sqrt{\vecb{C}^2 + m^2} - m \right) \epsilon^{ab} 
\lambda^a \lambda^b}{\vecb{C}^2} \ .
      \ee
Recalling (\ref{defslow}) and the { bosonic} ( for $k > 0$) nature of the ground state of the 
effective Hamiltonian (\ref{Heffsimplest}), we see that the ground states of the full Hamiltonian have the structure
 \be
 \Psi^{(0)}_{k>0} \ =\ \Phi_1(R_{jk}) \ + \ \Phi_2(R_{jk}) \, \epsilon^{abc}(A_1 - iA_2)^a 
\lambda^b \lambda^c
  \ee
  (where now $a = 1,2,3$). These wave functions are gauge invariant.
  \footnote{They are also invariant with respect to $O(2)$ rotations, see Eq.(\ref{psirotat}).}
  In the vicinity of the valley $\epsilon^{abc} A^b_j A^c_k = 0$ and for large $C \gg m$, the approximate equality
  $\Phi_1 \approx 4ig^2 |\vecb{C}|\Phi_2$  holds. Restoring the distinction between the fast and slow variables, 
we can represent
   \be
   \label{nearvalley}
   \Phi_1 \ =\ \Psi^{\rm slow}(C_j) \Psi^{\rm fast}_{\vecb{C}} (A^{1,2}_j)  \ ,
    \ee
    and the gauge invariance of $\Phi_1$, which means in particular its $G$-parity
    (invariance under rotations by $\pi$ along the second color axis), entails the Weyl invariance 
of $\Psi^{\rm slow}(C_j)$. We reproduce thereby our previous result.

  We are ready now to go over to the negative $k$ case and to understand how the paradox is resolved. 
 The point is that, when $k < 0$, the expression (\ref{PsivacfermanyC}) is {\it inconvenient}. 
The convenient expression is obtained from  (\ref{PsivacfermanyC}) by multiplying it by the factor $C_1 + iC_2$,
      \be
     \label{PsivacfermanyCnegk}
     \tilde{\Psi}^{\rm fast}_{\vecb{C}} \ =\
     4ig^2(C_1 + i C_2)   +  \left( \sqrt{\vecb{C}^2 + m^2} + |m| \right) \epsilon^{ab} \lambda^a \lambda^b \ .
      \ee
      Indeed, as far as the fast Hamiltonian and its eigenfunctions are concerned, the factors 
depending only on slow variables are absolutely irrelevant and can be chosen arbitrarily. 
The product $\psi \tilde \Psi$ can now be easily  promoted to a gauge-invariant eigenstate 
of the full Hamiltonian,
        \be
 \Psi^{(0)}_{k<0} \ =\ 2\Phi_2(R_{jk}) (A_1 + iA_2)^a \lambda^a \ + \ \frac 1{6} \Phi_1(R_{jk}) \, 
\epsilon^{abc}\lambda^a \lambda^b \lambda^c \ .
  \ee
 Again, $\Phi_1$ can be represented as in (\ref{nearvalley}), and the coefficients 
 $\Psi^{\rm slow}(C_j)$ (the  effective wave functions being obtained from them by multiplying by $\psi$) 
should be {\it even} rather than odd with respect to Weyl reflections, such that
  \be
  \label{indnegk}
  I^{\rm tree}(k < 0) \ =\ k-1 \ .
   \ee
   Going back to (\ref{PsivacfermanyCnegk}), one can notice that, in contrast to the function 
(\ref{PsivacfermanyC}), this function is {\it odd} with respect to rotations by $\pi$ around 
the second color axis producing the reflections $\vecb{C} \equiv \vecb{A}^3 \to - \vecb{C},\ \ 
\lambda^1 \to -\lambda^1, \lambda^2 \to \lambda^2$. This oddness compensates for the 
Weyl-oddness of the factor $\psi$ and requires for the coefficient $\Psi^{\rm slow}(C_j)$ to be Weyl-even.

 \vspace{2mm}

\centerline{\it Higher unitary groups.}

\vspace{2mm}

\begin{figure}[t]
\begin{center}
\includegraphics[width=3in]{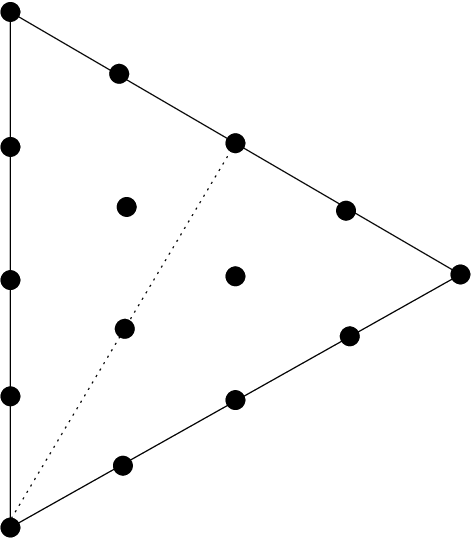}
\end{center}
\caption{$SU(3)$: 15 vacuum states for $k=4$. The dotted line marks the boundary of the Weyl alcove for $G_2$.}
\label{4k15}
\end{figure}

 Consider first $SU(3)$ and let $k$ be positive. There are $2r =4$ slow bosonic variables,
 which are convenient to choose as  $x^a = C_1^a L/(4\pi), \ \ y^a = C_2^a L/(4\pi)$. Both
$x^a$ and $y^a$ vary
 within an elementary cell of the $SU(3)$ coroot lattice, alias the maximal torus.
  The latter represents a rhombus
 shown in Fig.1 such that $\exp\{iLC^a t^a\} = 1$ in the vertices of the rhombus. The
  effective Hamiltonian (\ref{HeffN}) can be represented in the form
       \be
       \label{Heff3}
       H_{\rm eff} \ =\ \frac {g^2}{2L^2} \left(P_x^a - \frac {\kappa L^2}2 y^a \right)^2 +
  \frac {g^2}{2L^2} \left(P_y^a + \frac {\kappa L^2}2 x^a \right)^2 \ + \ \frac {\kappa g^2}{2} 
(\psi^a \bar\psi^a - \bar\psi^a \psi^a)\ .
        \ee
 We have found earlier [see Eq.(\ref{ISUN})] that this Hamiltonian has $3k^2$
 ground states. The corresponding wave functions represent generalized theta functions
 defined on the coroot lattice of $SU(3)$. %(see e.g. \cite{Serr}). 
They satisfy the boundary conditions
     \be
     \label{bc3}
     \Psi(\vecb{x}+\vecb{a}, \vecb{y}) &=& e^{-2\pi i k\vecb{a} \vecb{y}} \Psi(\vecb{x},\vecb{y}) \ , \nonumber \\
       \Psi(\vecb{x}+\vecb{b}, \vecb{y}) &=& e^{-2\pi ik \vecb{b} \vecb{y}} \Psi(\vecb{x},\vecb{y}) \ ,
       \nonumber \\
     \Psi(\vecb{x}, \vecb{y}+\vecb{a}) &=& e^{2\pi i k \vecb{a} \vecb{x}} \Psi(\vecb{x},\vecb{y}) \ ,
        \nonumber \\
      \Psi(\vecb{x}, \vecb{y}+\vecb{b}) &=& e^{2\pi i k \vecb{b} \vecb{x}} \Psi(\vecb{x},\vecb{y}) \ ,
      \ee
  where     $\vecb{a} = (1,0), \ \vecb{b} = (-1/2, \sqrt{3}/2)$ are simple coroots.
 When  $k = 1$, there are 3 such states:
  \be
  \label{psi3k1}
  \Psi_0 &=& \sum_{\vecb{n}} \exp \left\{ -2\pi (\vecb{n} + \vecb{y})^2 -
 2\pi i \vecb{x} \vecb{y} - 4\pi i \vecb{x} \vecb{n} \right \} \ , \nonumber   \\
  \Psi_\triangle &=& \sum_{\vecb{n}} \exp \left\{ -2\pi (\vecb{n} + \vecb{y} +
  {\triangle\!\!\!\!\!\triangle})^2 - 2\pi i \vecb{x}
  \vecb{y} - 4\pi i \vecb{x} (\vecb{n} + {\triangle\!\!\!\!\!\triangle}) \right \} \ , \nonumber    \\                    
      \Psi_\Box &=& \sum_{\vecb{n}} \exp \left\{ -2\pi (\vecb{n} + \vecb{y} +
    {\Box\!\!\!\!\!\Box})^2 - 2\pi i \vecb{x}
        \vecb{y} - 4\pi i \vecb{x} (\vecb{n} + {\Box\!\!\!\!\!\Box} )\right \}\ ,           
   \ee
 where the sums run over the coroot lattice,  $\vecb{n} = m_a \vecb{a} + m_b \vecb{b}$
 with integer $m_{a,b}$.
  Now, ${\triangle\!\!\!\!\!\triangle}, \ {\Box\!\!\!\!\!\Box}$ are certain special points on the maximal torus
  (called {\it fundamental coweights}) satisfying
  $${\triangle\!\!\!\!\!\triangle} \vecb{a} =
  {\Box\!\!\!\!\!\Box} \vecb{b} = 1/2, \ \ \ \   {\Box\!\!\!\!\!\Box} \vecb{a} =
                                      {\triangle\!\!\!\!\!\triangle} \vecb{b} = 0 \ .$$
The group elements that correspond to the points $0, \triangle$, and $\Box$ belong to the center of the group, 
 \be
\label{center}
U_0 &=& {\rm diag} (1,1,1) \ , \nonumber \\
 U_\Box &=& {\rm diag}  (e^{2i\pi/3}, e^{2i\pi/3}, e^{2i\pi/3})\ , \nonumber \\
 U_\triangle &=& {\rm diag}(e^{4i\pi/3}, e^{4i\pi/3}, e^{4i\pi/3}) \ .
 \ee
They are obviously invariant with respect to Weyl symmetry, which permutes the eigenvalues.
\footnote{\label{granicy} For a generic coweight, the Weyl group elements permuting the elements $(12)$, $(13)$ and
$(23)$ are represented as the reflections with respect to the dashed lines bounding the Weyl 
alcove ($\equiv$ the quotient $T/W$) in Fig. \ref{romb}.} Thus, all three states (\ref{psi3k1}) at the level $k=1$ are Weyl invariant. But for  
 $k > 1$, the number of invariant states is less than $3k^2$.  
For an arbitrary $k$ and in the constant field, the wave functions of all $3k^2$ eigenstates
can be written in the same way as in (\ref{psi3k1}),
   \be
   \label{psi3anyk}
\Psi_n \ =\ \sum_{\vecb{n}} \exp \left\{ -2\pi (\vecb{n} + \vecb{y} +
    \vecb{w}_n)^2 - 2\pi i \vecb{x}
        \vecb{y} - 4\pi i \vecb{x} (\vecb{n} + \vecb{w}_n )\right \}\ , 
   \ee
where $\vecb{w}_n$ are coweights whose projections on the simple coroots $\vecb{a}, \ \vecb{b}$ represent
 integer multiples of $1/(2k)$. Only the functions (\ref{psi3k1}) with $\vecb{w}_n$ lying of the vertices of the Weyl alcove
are Weyl invariant. For all  other $\vecb{w}_n$, one should construct Weyl invariant combinations
   \be
\label{sumWeyl}
\Psi \ =\ \sum_{\hat{x} \in W} \hat{x} \Psi_{\vecb{w}_n} \ .
 \ee 
 As a result, the number of Weyl invariant states is equal to the number 
of the coweights $\vecb{w}_n$ lying within the Weyl alcove.
   For example, in the case $k=4$, there are
     15 such coweights shown in Fig.\ref{4k15} and, correspondingly, 
15 vacuum states.

       For a generic $k$, the number of the states is
 \be
I^{\rm tree}_{SU(3)}(k>0) \ =\  \sum_{m=1}^{k+1} m     \ =\ \frac {(k+1)(k+2)}2  \ .
  \ee

  The analysis for  $SU(4)$ is similar. The Weyl alcove is the tetrahedron  with the vertices
corresponding to the center elements of $SU(4)$. A pure geometric counting gives 
                  \be
               I^{\rm tree}_{SU(4)}(k>0) \ =\  \sum_{m=1}^{k+1} \sum_{p=1}^m p   \ =\ \frac {(k+1)(k+2)(k+3)}6  \ .
                   \ee
   The generalization for  an arbitrary $N$ is obvious. It gives the result (\ref{nashindextree}).

          The large $k$ asymptotics is $I \sim k^{N-1}/(N-1)!$, which is simply the "pre-Weyl"
          index (\ref{ISUN}) divided by the order of the Weyl group.
          For negative $k$, the ground states of the effective Hamiltonian acquire
       the fermionic  factor
       $$ \sim \prod_{a=1}^{N-1} \psi^a \ . $$
        For odd $N$, the ground states
       are still bosonic and the index is still positive. For even $N$,
       the ground states are fermionic and the index is negative.
       One need not perform here a detailed analysis, as we did in the case of $SU(2)$, but simply 
use the symmetry requirements. They dictate the formula
           \be
\label{ItreeSUN}
               I^{\rm tree}_{SU(N)}({\rm any}\ k) \ =\ [{\rm sgn}(k)]^{(N-1)}
                \left( \begin{array}{c}  N+|k|-1 \\ N-1 \end{array}
               \right) \ .
                   \ee

    \vspace{2mm}

    \centerline{\it Symplectic groups}

             \vspace{2mm}

   The counting of vacuum states for the symplectic groups $Sp(2r)$ is
   simpler (sympler ?)  than for unitary groups. The maximal torus of $Sp(2r)$ can be represented as
   $g = \exp\left\{i \sum_{p=1}^r \alpha_p e_p \right\}$, where
   \be
   \label{ek}
   e_1 &=& \frac 12 {\rm diag}\, (1,0,\ldots,0,-1) \nonumber \\
       & \cdots &    \nonumber \\
   e_r &=& \frac 12 {\rm diag}\, (0,\ldots,0,1,-1,0,\ldots,0)
    \ee
    is the orthonormal basis in the Cartan subalgebra and $\alpha_k \in (0,4\pi)$.
    The coroot lattice
    is thus hypercubic.
\footnote{ There are $r-1$ long and one short  simple coroot,
    $$ a_1 = e_1-e_2,\ \ \ldots, \ \ a_{r-1} = e_{r-1} - e_r, \ \
    b = e_r \ .$$
But the basis (\ref{ek}) is more convenient than the basis $\{a_1,\ldots,a_{r-1}, b\}$.  }

The effective BO Hamiltonian represents a simple sum of $r$ copies of the BO Hamiltonian for
$Sp(2) \equiv SU(2)$. The path integral for the pre-Weyl index is the $r$-th power
of such path integral for $SU(2)$ giving
  \be
  I^{Sp(2r)}_{\rm pre-Weyl} \ =\ (2k)^r \ .
   \ee
   The vacuum wave functions represent the products of the $SU(2)$ wave functions
   (\ref{Psim}).
   The Weyl group changes the sign for each $\alpha_k$ and permutes them. Its order is thus
   $2^r r!$. Thus, the number of Weyl-invariant vacuum states can be counted as the number
   of components of a symmetric tensor of rank $r$ where each index can take $k+1$ values.
   For positive $k$, it is given by Eq.(\ref{indSp}). The index for negative $k$ is restored
   by symmetry, $I(-k) = (-1)^r I(k)$.

\vspace{2mm}

    \centerline{$G_2$.}

             \vspace{2mm}

\begin{figure}[t]
\begin{center}
\includegraphics[width=4in]{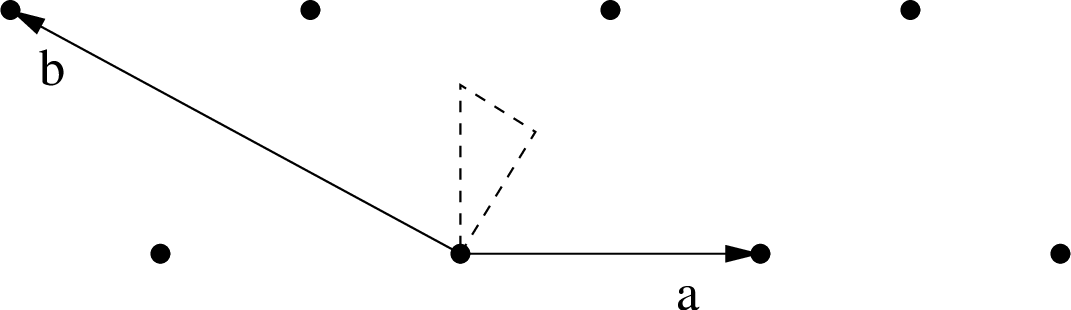}
\end{center}
\caption{Coroot lattice and Weyl alcove for $G_2$.}
\label{G2stuff}
\end{figure}

The simple coroots for $G_2$ are $\vecb{a} = (1,0)$ and $\vecb{b} = (-3/2, \sqrt{3}/2)$. The lattice of coroots
and the maximal torus look exactly in the same way as for $SU(3)$ (see Fig. \ref{G2stuff}). Hence, the pre-Weyl index
is equal to $3k^2$, as for $SU(3)$. The difference
is that the Weyl group involves now 12 rather than 6 elements, and the Weyl alcove is two times smaller than for
$SU(3)$.   As a result, for $k=4$, we have only 9 (rather than 15) Weyl-invariant states (see Fig.\ref{4k15}).
The general formula is given in Eq.(\ref{indG2}).

\section{Loop corrections.}

   \setcounter{equation}0

  In (nonchiral) $4d$ SYM theories, the evaluations of the index based on the analysis
  of the tree effective BO Hamiltonian are not modified when loops are taken into account.
  For $3D$ SYMCS theories, this is not so and loop effects are important. It seems
  plausible, however, that one can restrict oneself by
  {\it one-loop} analysis; second and higher loops do not further modify the result. We will argue
this point a bit later. 

  \subsection{Infinite volume.}
 We are interested in one-loop corrections to the effective Hamiltonian in finite volume.
 But they are genetically related to one-loop renormalization of the infinite volume theory
 \cite{sestry}.
 For pure YMCS theory, the latter was dealt with in Ref.\cite{Rao}. For ${\cal N} = 1,2,3$
  SYMCS theories, the corresponding calculations
  have been performed in \cite{Kao}. Let us remind their salient features.

  After fixing the gauge and introducing the ghosts, the Lagrangian acquires the form
  \be
  \label{Lwithghosts}
   {\cal L} \ =\ - \frac 1{2g^2} {\rm Tr} \left \{ F_{\mu\nu}^2 \right \} +
   \kappa   \epsilon^{\mu\nu\rho} {\rm Tr}
   \left\{ A_\mu \partial_\nu A_\rho - \frac {2i}3 A_\mu A_\nu A_\rho \right \} + 
\partial^\mu \bar \eta^a \partial_\mu \eta^a \nonumber \\
   + f^{abc} \partial^\mu \bar \eta^a A_\mu^b \eta^c -
   \frac 1{2\xi} (\partial^\mu A_\mu^a)^2\ \ + \ {\rm fermion\ terms}                                                                   
   \ee
   It is convenient to use Landau gauge $\xi \to 0$. Then the tree gluon propagator is
   \be
   \label{Del0}
   \Delta_{\mu\nu}^{(0)} \ =\ - \frac {ig^2}{p^2 - m^2} \left[ g_{\mu\nu} -
   \frac {p_\mu p_\nu}{p^2} + \frac {im \epsilon_{\mu\nu\rho} p^\rho}{p^2} \right] \ .
    \ee
    It has the pole at $p^2 = 0$ associated with gauge degrees of freedom and the physical pole
    at $p^2 = m^2$.  The transverse gluon polarization operator has two structures
     \be
     \label{Pimunu}
     \Pi_{\mu\nu} \ =\ (p^2 g_{\mu\nu} - p_\mu p_\nu ) \Pi_e(p^2) - i \epsilon_{\mu\nu\rho}
     p^\rho \Pi_o(p^2) \ .
      \ee
      Introducing also the ghost polarization operator $\tilde \Pi(p^2)$,
      the  bosonic part of the renormalized Lagrangian is expressed as
      \footnote{It happens that the ghost-ghost-gluon vertex is not
      renormalized in Landau gauge at the one-loop level.}
      \be
      \label{Lren}
      {\cal L}^{\rm ren} \ =\
 &-& \frac 1{2g^2}[1 + g^2 \Pi_e(0)]  {\rm Tr} \left \{ F_{\mu\nu}^2 \right \} +
\kappa \left( 1 - \frac {\Pi_o(0)}\kappa \right)  \epsilon^{\mu\nu\rho} {\rm Tr}
\left\{ A_\mu \partial_\nu A_\rho  - \frac {2i}3 A_\mu A_\nu A_\rho \right \}
\nonumber \\
&+& [1 + \tilde \Pi(0)]
\partial^\mu \bar \eta^a \partial_\mu
\eta^a
+ f^{abc} \partial^\mu \bar \eta^a A_\mu^b \eta^c + \ \ \ {\rm gauge\ fixing\ term}
\ .
\ee
($\tilde \Pi(0)$ is the ghost polarization operator).  Redefining the fields $\eta, A$, it can be rewritten as
   \be
   \label{Lredef}
        {\cal L}^{\rm ren} \ =\
   - \frac 1{2g_{\rm ren}^2}  {\rm Tr} \left \{ F_{\mu\nu}^2 \right \} +
 \kappa_{\rm ren}  \epsilon^{\mu\nu\rho} {\rm Tr}
  \left\{ A_\mu \partial_\nu A_\rho + \dots \right \}
+ \partial_\mu \bar \eta^a \partial_\mu \eta^a + \nonumber \\ f^{abc} \partial_\mu \bar \eta^a A_\mu^b \eta^c + \ldots   ,
  \ee
 where
 \be
 \label{kapgren}
 \kappa_{\rm ren} \ =\ \kappa \left[ 1 - \frac 1\kappa \Pi_o(0) + 2 \tilde \Pi(0) \right]
  \ , \nonumber \\
  \frac 1{g^2_{\rm ren}} \ =\ \frac 1{g^2} \left[ 1 + g^2 \Pi_e(0) + 2 \tilde \Pi(0) \right]  \ .
   \ee

\begin{figure}[t]
\begin{center}
\includegraphics[width=5in]{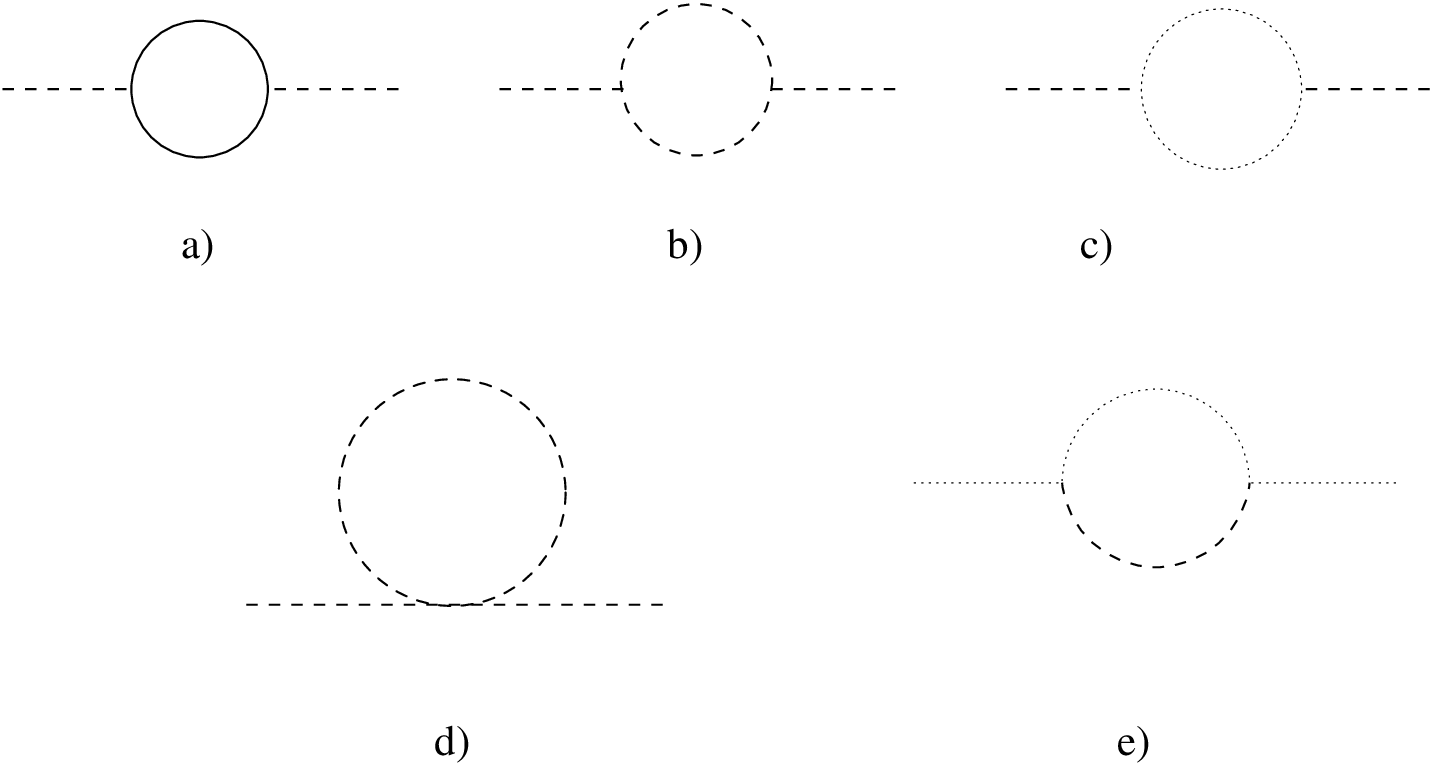}
\end{center}
\caption{Gluon and ghost polarization operators.}
\label{graphsCS}
\end{figure}

 The relevant 1-loop graphs are depicted in Fig.\ref{graphsCS}.
   Let us discuss first the renormalization of $\kappa$. The simplest is the
   contribution of the fermion loop in Fig.\ref{graphsCS}a. It gives
    \be
    \label{PioF}
    \Pi^F_o(0) \ =\ mc_V \int \frac {d^3p}{(2\pi)^3} \frac 1{(p^2 + m^2)^2} \ =\
    \frac {c_V}{8\pi}\ ,
     \ee
    $p$ being the Euclidean momentum. The bosonic contribution can be obtained from Eqs.(17,18) of Ref.\cite{Kao},
     \be
     \label{PioBiPitilda}
     \Pi^B_o(0) - 2\kappa \tilde \Pi(0) \ =\
     - \frac {2mc_V }3 \int  \, \frac {d^3p}{(2\pi)^3}  \frac {5p^2 + 2m^2}{p^2(p^2 + m^2)^2}
     +    \frac {4m c_V }3 \int \, \frac {d^3p}{(2\pi)^3}  \frac {1}{p^2(p^2 + m^2)} \nonumber \\
      =  - 2 mc_V \int \, \frac {d^3p}{(2\pi)^3} \frac 1{(p^2 + m^2)^2} = -\frac {c_V}{4\pi} \ .
      \ee
      Thus, the fermion loop leads to the shift $k \to k - c_V/2$, while the boson loops
      give $k \to k+ c_V$, such that 
  \be
\label{kren}
k_{\rm ren} \ = \ k + c_V/2\ .
 \ee
 (That was for positive $k$.
      When $k$ is negative, $k_{\rm ren} = k - c_V/2$. ) Note the same structure
      of momentum integrals in the fermionic and bosonic contributions.

      The coefficient $k$ is not renormalized beyond one loop. The proof is simple.
      Consider the case  $k \gg c_V$. This is the perturbative regime where the loop corrections
      are ordered such that $\Delta k^{\rm (1\ loop)} \sim O(1),\ \  \Delta k^{\rm (2\ loops)} \sim O(1/k)$, etc. 
But the corrections of order $\sim 1/k$ to $k$ are not allowed. To provide for
      gauge invariance, $k_{\rm ren}$ should be integer. Thereby, all higher loop
      contributions in $k_{\rm ren}$ must vanish.

When $c_V$ is odd (in particular, when $N$ is odd for $SU(N)$ groups), the coefficient $k$ is shifted by a 
half-integer. The physical requirement for $k$ to be integer refers to $k_{\rm ren}$ rather than $k_{\rm tree}$. 
This implies that, for consistency, $k_{\rm tree}$ should be half-integer.
\footnote{Another way to see this is to notice that, for odd $c_V$, a topologically nontrivial gauge transformation
brings about the extra factor $-1$ due to the level flow in the fermion determinant. As a result, the quantization
condition is not $\exp\{2\pi i k_{\rm tree} \} = 1$, but rather $\exp\{2\pi i k_{\rm tree} \} = -1$, giving
half-integer  $k_{\rm tree}$  \cite{Redlich,Wit99}. }

      The renormalization of the coefficient $1/g^2$ of the kinetic term can be obtained from
the result 
 \be
\label{massren}
m_{\rm ren} \ =\ m \left[ 1 + \frac {2c_V}k \right]
 \ee 
for the mass renormalization  
\footnote{See Eq.(23) in Ref.\cite{Kao}. Note that Eq.(22) there involves a misprint with misplaced factor $\ln 3$.}, 
from Eq.(\ref{kren}), and from the relation
(\ref{mass}). One obtains
  \be
\label{g2ren}
  \frac 1{g^2_{\rm ren}}  \ =\ \frac 1{g^2} \left( 1 - \frac {3c_V}{2k} \right)        \ .
 \ee
    
\vspace{2mm}

\centerline{\it Background field calculation.}

\vspace{2mm}

The calculations \cite{Rao,Kao} were done in the conventional diagrammatic approach. But to generalize them to the finite volume case, 
the background field technique is more appropriate and relevant. We are not aware of a honest background field calculation
in SYMCS or YMCS systems. However, the bosonic shift $ k \to k + c_V$ can be reproduced rather easily in the background field technique,
if making a little surgery in the regulator sector and replacing the gauge-invariant YM action by a simple-minded gluon mass term 
\cite{WitCMP,Shifman}. Consider the pure CS term and split the gauge field $A_\mu$ in two parts,
 \be
\label{split}
 A_\mu \ =\ A_\mu^{\rm cl} +  \frac 1{\sqrt{\kappa}}  a_\mu^{\rm qu}\ 
 \ee  
(the factor $1/\sqrt{\kappa}$ being introduced for convenience). 
The background  field $A_\mu^{\rm cl}$ is assumed to satisfy the classical equations of motion $F^{\rm cl}_{\mu\nu} = 0$.
Then the CS action is reduced in the quadratic in $a$ approximation to 
 \be
\label{actionsplit}
 S_{\rm CS} \ =\  S_{\rm CS} (A_\mu^{\rm cl}) +  \epsilon^{\mu\nu\rho} \, {\rm Tr} \int d^3x \, a_\mu D_\nu a_\rho \ .
 \ee     
To do perturbative calculations, one has to fix the gauge. The most convenient one is the 
background Landau gauge $D^\mu a_\mu = 0$, where the covariant derivative $D_\mu$ involves only the classical part.
 We are using then a slightly nonstandard way to implement this gauge
condition by introducing the Lagrange multiplier $\phi$ and adding to the Lagrangian the term 
 \be
\label{Lagrmultip}
\Delta {\cal L} \ =\  \left[ \phi D^\mu a_\mu - (D^\mu \phi) a_\mu \right] \ .
 \ee
There are also ghosts with the Lagrangian ${\cal L}_{\rm ghost} = -{\rm Tr} \left\{ \bar c D_\mu^2 c \right\} $, but they
do not affect the renormalisation of $\kappa$ we are interested in. One can now combine $a_\mu$ and $\phi$ into a four-dimensional
object $B_M = \{a_\mu, \phi\}$ \, ($M = 1,\ldots,4$), such that the (relevant part of) the quantum action takes form
 \be
\label{Squ}
S_{\rm qu} \ =\ i  {\rm Tr} \int d^3x  \, B D^\mu \Gamma_\mu B \ , 
 \ee 
  where $\Gamma_\mu$ are certain traceless $4 \times 4$ matrices satisfying the same (anti)commutation relations as Pauli matrices.
This Lagrangian is very similar to the Dirac Lagrangian. There are two differences: {\it (i)} There are twice as many $B_M$'s as $\lambda_\alpha$'s,
giving a twice as large contribution to the effective action. {\it (ii)} $B_M$ are bosons rather than fermions and contribute to the effective action 
with an opposite sign. 

Thus, the bosonic contribution in this approach has exactly the same structure as the fermion one, up to a factor $-2$. 
Of course, we are a little bit cheating here. The renormalization of $\kappa$ in the theory with the action (\ref{Squ}) is 
zero or, better to say, not defined until it is regularized in the infrared. A natural regularization is provided by 
the Yang-Mills term in the action. But the calculation of Ref.\cite{Shifman} uses instead a simple-minded regularization consisting in 
adding to the Lagrangian the gauge boson mass term 
 \be
\label{massterm}
\Delta {\cal L}_m \ \sim \ m{\rm Tr} \{B^M B_M \}
 \ee
 (with a properly chosen sign). This regularization is not so nice
as the YM one (it is not gauge invariant, etc), but it has the advantage that the calculations become very simple. 
Actually, one does not need to   do them again, but can simply use the fermion results. This gives $\Delta k_{\rm bos} = 
-2\Delta k_{\rm ferm} = c_V$, which coincides with the result of \cite{Rao}.

        \subsection{Finite volume.}

  Consider first the $SU(2)$ theory.
 As was mentioned above, the coefficient $\kappa$ (with the factor $L^2$) has the meaning
  of magnetic field on the dual torus for the effective finite volume BO Hamiltonian.
  Renormalization of $\kappa$ means renormalization of this magnetic field. At the tree level,
   the magnetic field was constant. The renormalized field is not constant, but depends
   on the slow variables $\vecb{C}$. To find this dependence, one has to substitute 
   \be
   \label{pnC}
   \vecb{p} &\to & 2\pi \vecb{n}/L - \vecb{C} \nonumber \\
 \int \frac {d\vecb{p} }{(2\pi)^2} & \to & \frac 1{L^2} \sum_{\vecb{n}}    
\ee
    in the integral
   $\sim \int d^3p$ for $\Delta \kappa$. 
\footnote{Heuristically, the appearance of $\vecb{C}$ in (\ref{pnC}) is due to replacing the usual derivative
by the covariant one, evaluated in the constant potential background. The accurate calculation performed for the fermions 
in the Appendix
confirms this rule, but we will see that the mechanism by which the combination $2\pi\vecb{n}/L - \vecb{C}$ arises is not so trivial.
It appears, indeed, in the integral for the induced magnetic field, but the corresponding integral for
${\cal A}_k(\vecb{C})$ (which enters the effective Lagrangian) has a more complicated structure.} 
 We derive for  positive $k$ 
   \be
   \label{DBferm}
    \Delta {\cal B}^F(\vecb{C}) &=& 
   -2m  \int_{-\infty}^\infty \frac {d\epsilon}{2\pi} \sum_{\vecb{n}}
   \frac 1{\left[\epsilon^2 + \left( \frac {2\pi \vecb{n}}L - \vecb{C} \right)^2 +
   m^2 \right]^2}  \nonumber \\ 
   = - \frac {m }2 \sum_{\vecb{n}}
   \frac 1{\left[\left( \frac {2\pi \vecb{n}}L - \vecb{C} \right)^2 +
   m^2 \right]^{3/2}} \ , \nonumber \\ 
   \Delta {\cal B}^B(\vecb{C}) &=& -2 \Delta {\cal B}^F(\vecb{C})\ .
\ee

For most values of ${\vecb C}$, this correction is of order $\sim mL^3 = \kappa g^2 L^3$,
 which is small compared to ${\cal B}^{\rm tree} \sim \kappa L^2$ if $g^2 L \ll 1$,
 which we assume. Also in the "corner" of the torus $|{\vecb C}| \ll m$, the correction $\Delta {\cal B} \sim 1/m^2$ is 
small compared to  ${\cal B}^{\rm tree}$ for very large $k$, $ k \gg 1/(mL)^2$. Otherwise, 
 $\Delta {\cal B}$ dominates there. 
 \footnote{\label{fixed} The same concerns all the points $\vecb{C} = 2\pi\vecb{n}/L$.
Note that there are {\it four} such points in the dual torus $ C_j \in (0, 4\pi/L)$ with
$\vecb{n} = (0,0), \ (0,1), \ (1,0), \ (1,1)$,  corresponding to the group elements $\Omega_j = \exp\{iL t^3 C_j \} = \{\pm 1, \pm1\}$.
 These points are invariant with respect to the action of the Weyl group.}
In any case, the integral for the {\it flux}  associated with the 
 corrections (\ref{DBferm}) is saturated by the regions $|{\vecb C}| \lsim m$, etc in the vicinity of Weyl fixed points,
  being equal to
  \be
  \label{counterflux}
  \Delta \Phi^F \ =\ - \frac {m}2 \sum_{\vecb{n}} \int_{C_j \in (0, 4\pi/L)}
  d^2C \frac 1{\left[\left( \frac {2\pi \vecb{n}}L -
  \vecb{C} \right)^2 +
       m^2 \right]^{3/2}} \nonumber \\
  = - 2m \int \frac {d^2p}{(p^2 + m^2)^{3/2}} \ = \ - 4\pi  \ ,
  \ee
     which should be compared with the tree flux $\Phi^{\rm tree} = 4\pi k$. The total flux is
     thus
     \be
     \label{fluxtot}
     \Phi^{\rm tot} \ = \ \Phi^{\rm tree} + \Delta \Phi^F + \Delta \Phi^B =
     4\pi \left( k - 1 + 2  \right)
     \ee
     The  renormalized flux means the renormalized index. For $SU(2)$, we obtain
     the result $2(k+1)$ for the pre-Weyl index. After taking into account the Weyl invariance condition, we
derive
 \be
\label{ItotN2}
I(k\neq 0) \ =\ {\rm sgn}(k) (|k|+2) \ .
 \ee  
 When $k=0$, the  magnetic flux  giving
the pre-Weyl index 
 is zero, and loop corrections do not modify 
this result (when $k_{\rm tree}$ vanishes, this is also the case for $k_{\rm ren}$). A vanishing index suggests breaking of 
supersymmetry, but whether or not supersymmetry is actually broken in this case 
is a nontrivial question requiring special
studies.

The result (\ref{ItotN2}) involves the tree contribution and the one-loop correction. One can argue that higher-loop corrections must
vanish. The reasoning is the same as for renormalization of $k$ in the infinite volume: for large $k$ a two-loop correction should
be suppressed as $\sim 1/k$. But the coefficient of $1/k$ should vanish - otherwise the renormalized flux and renormalized index 
would not be integer.  

A similar analysis (see Appendix) can be done for the groups of higher rank. It displays that, at the level of one loop, 
the generalized magnetic flux
(\ref{indCecotti3}) evaluated with renormalized ${\cal B}^{ab}(\vecb{C}^a)$ is obtained from the corresponding tree expression
by substituting $k \to k + c_V/2$. This suggests (though does not prove rigourously ) that there is no nontrivial renormalization of the
generalized flux due to second and higher loops.
\footnote{\label{nohigherloop} To prove it, one has to exclude a nontrivial two- and higher-loop renormalization of the generalized flux 
not reduced to renormalization
of $k$.}
   For $SU(N>2)$, the result is 
  \be
  \label{nashindexpmk}
  I (k\neq 0,N)  \ = \  [{\rm sgn}(k)]^{N-1} \left( \begin{array}{c} |k| + 3N/2 -1 \\ N-1 \end{array} \right)  \ .
  \ee
For symplectic groups,
  \be
 \label{indSppmk}
 I(k\neq 0,r)  \ =\ [{\rm sgn}(k)]^r  \left( \begin{array}{c} |k| + 3r/2 + 1/2 \\ r \end{array} \right)\ .
  \ee
For $G_2$, \ $c_V=4$, and the result for the index is given by the expression (\ref{indG2}) with $|k|$ being substituted by
$|k| + 2$. 
When $k=0$ (this is allowed for even $N$ and for odd $r$), the index vanishes.

      \vspace{1mm}

      \centerline{\it Metric and the index.}

         \vspace{2mm}
Let us restrict ourselves here by the discussion of $SU(2)$. 
The index is a topological quantity and is determined by relevant  topological invariants, like
the magnetic flux (alias, the first Chern class of the relevant to the problem $U(1)$ bundle on the 
moduli space of flat $SU(2)$ connections on $T^2$). 
\footnote{The pre-Weyl index is just equal to the first Chern class. The index taking into account the Weyl-invariance of wave functions
can  be related to a certain more complicated invariant \cite{Wit99}.}
Thus, it is sensitive only to the  modifications of the flux
due to loops and is {\it robust} with respect to other loop corrections to BO Hamiltonian. 
In particular, the index is not sensitive to corrections to the {\it metric}, which are well there and might modify 
significantly the effective Hamiltonian in corner of the torus and other Weyl fixed points.  
\footnote{
Such corrections are also present for $4d$ theories.
In $4d$ supersymmetric QED, they
       were calculated by us long time ago \cite{ja87}. For non-Abelian $4d$ theories, this was done
       in \cite{ja02}.}
At the one loop level, these corrections are associated to the renormalization (\ref{g2ren}) 
of the coupling $1/g^2$ by the same token as the correction
(\ref{DBferm}) is associated with renormalization of $\kappa$. The explicit calculation gives
 \be
\label{corrmetr}
\delta g^{\rm 1\ loop}(\vecb{C}) \ =\ \frac {c_V g^2}{ L^2 \left(\vecb{C}^2 + m^2 \right)^{3/2}} \ .
 \ee 
To see insensitivity of the index to the metric explicitly, let us write the supersymmetric Hamiltonian for the system with 
nontrivial metric and calculate the corresponding
phase space integral, as in Eq.(\ref{indCecotti}). The Hamiltonian is derived from the supersymmetric Lagrangian
  \be
\label{Lmetrsuper}
{\cal L}\ =\ \int d \bar\theta d\theta  \left[ \frac {g(\bar Z, Z)}4 {\cal D} \bar Z \bar {\cal D} Z + \Phi( \bar Z, Z) \right] \ ,
 \ee
with 
\be
\label{covD}
 {\cal D} \ =\ - \frac \partial {\partial \bar\theta } + i\theta \frac \partial {\partial t}, 
\ \ \ \ \ \ \ \bar{\cal D} \ =\  \frac \partial {\partial \theta } - i \bar \theta \frac \partial {\partial t} \ .
 \ee 
$Z$ is a chiral superfield, ${\cal D} Z = 0$, which is convenient to write in components as
\be
\label{Z}
Z= z +  \frac {\sqrt{2}} {\sqrt{ g(\bar z, z)}} \theta \bar\psi  - i\dot z \theta \bar\theta \ .
 \ee
 Then
\be
\label{Lmetric} 
{\cal L} \ =\ g \dot{\bar z} \dot {z} - \dot{z}   {\cal A}  -  \dot{\bar z} \bar {\cal A} 
 + \frac i2 (\bar \psi {\dot \psi} - \dot {\bar\psi} \psi )
+ \frac {i \bar\psi \psi }2 \left( \dot {\bar z} \bar \partial \ln g - {\dot z} \partial \ln g \right)
+ \frac {{\cal B}}{g}  \bar\psi \psi \ ,
 \ee
where  ${\cal A} = i\partial \Phi$ and ${\cal B} = 2\partial \bar\partial \Phi$. The canonical Hamiltonian is
          \be
        \label{Hammetric}
        H \ =\ f^2  (\pi + {\cal A})(\bar \pi + \bar {\cal A}) + 
 i f \bar \psi \psi \left[ \pi \bar \partial f 
- \bar\pi \partial f 
        +  \bar \partial ({\cal A} f ) - \partial (\bar {\cal A} f ) \right] 
         \ee
with $f = g^{-1/2}$. 
It can be represented as the Poisson bracket $\{\bar Q, Q \}$ of the supercharges
          \be
        \label{Qmetric}
      Q &=& f(z, \bar z) \psi (\pi + {\cal A} ) \nonumber \\
      \bar Q &=&   f(z, \bar z) \bar \psi (\bar \pi + \bar {\cal A} ) \ .
       \ee
When $f=1$, the Hamiltonian (\ref{Hammetric}) coincides with (\ref{HeffN}) (with $g^2/L^2$ set  to 1 and  
color index $a$ suppressed) 
after identification
 $z = (C_1 - iC_2)/\sqrt{2}, \ \pi = (P_1 + iP_2)/\sqrt{2}, {\cal A} = ({\cal A}_1 + i{\cal A}_2)/\sqrt{2}$.
It is straightforward to see that the index  does not depend on the metric and the relation (\ref{indCecotti}) still holds.

\subsection{Higher ${\cal N}$.}

\vspace{2mm}

\centerline{${\cal N} =2$}

\vspace{2mm}

The ${\cal N} = 2$ SYMCS theory involves one more adjoint Majorana fermion and an extra adjoint real scalar $\Phi$. 
It has the following Lagrangian
  \be
 \label{LN2}
  {\cal L} \ =\ \frac 1{g^2} {\rm Tr} \int d^3x \left \{ - \frac 12 F_{\mu\nu}^2 + (D_\mu \Phi)^2 - m^2 \Phi^2 + 
  i\bar \lambda_f /\!\!\!\!D \lambda_f \right \} + \nonumber \\
  \kappa {\rm Tr}\int d^3x  \left\{ \epsilon^{\mu\nu\rho}
  \left( A_\mu \partial_\nu A_\rho - \frac {2i}3 A_\mu A_\nu A_\rho \right ) - \bar \lambda_f \lambda_f \right \} \ ,
   \ee
$f = 1,2$.
Its  Yang-Mills part  
is obtained by dimensional reduction from the standard ${\cal N} = 1$ $4d$ SYM theory.  The {effective} finite-volume  Lagrangian 
depends now on $3r$ bosonic  variables $\vecb{C}^a,  \Phi^a$, and on $2r$ holomorphic fermion variables $\lambda_f^a$, $a = 1,\ldots,r$.  
The  Lagrangian enjoys ${\cal N} = 2$ SQM symmetry.
\footnote{This is in a conservative convention meaning that the system involves two
{\it complex} conserved supercharges. Most experts in SQM prefer now to define ${\cal N}$ as the number of 
real supercharges, but, bearing in mind that
the theory with a single real supercharge is in fact {\it not} supersymmetric (does not involve a double spectral degeneracy), 
this devaluation of ${\cal N}$ does not seem to us 
to be a convenient innovation.}

 Similar to the effective Lagrangian for {\it chiral} $4d$ theories \cite{ZhETF,Blok}, the Lagrangian belongs to the class of generalized 
de Crombrugghe--Rittenberg supersymmetric Hamiltonians \cite{CromRit}. When $r=1$, the latter reads
  \be
\label{RitCrom}
H \ =\ \frac 12 (\vecb{P} + \vecb{{\cal A}} )^2 + \frac 12 K^2 + {\cal B}_j \bar \psi \sigma_j \psi \ ,
 \ee
where $\vecb{{\cal B}} = \vecb{\nabla} \times \vecb{\cal A} = - \vecb{\nabla} K$, and  
$K$ is an arbitrary function of three bosonic variables $\vecb{A}$.
For chiral $4d$ QED, the function $K$ was singular, $K \propto 1/|\vecb{A}|$. The corresponding Hamiltonian described the motion
in a monopole  field with extra scalar potential $\sim 1/\vecb{A}^2$. The singularity at $|\vecb{A}|=0$ 
led to nontrivial Berry's phase \cite{ZhETF}. 
In our case, $K$ is much simpler, $K = m\tilde \Phi$ ($\tilde \Phi \equiv \Phi^{3(\vecb{0})}$).
This corresponds to  uniform  magnetic field supplemented by an oscillatoric potential in $z$ direction. 
The Hamiltonian can be presented in the form
  \be
    \label{HeffN2}
   \frac {L^2}{g^2} H \ =\ \frac 1{2} \left(P_j - \frac {B}2 \epsilon_{jk} C_k \right)^2 + \frac {\tilde P^2
+  B^2 \tilde \Phi^2}2
 + \frac {B}2  (\psi_f \bar\psi_f - \bar\psi_f \psi_f)
    \ee
with $B = \kappa L^2$ (to establish its relationship to (\ref{RitCrom}), one should rename $\bar \psi_2 \leftrightarrow \psi_2$). 
In spite of the presence of potential (such that the configuration space $C_j, \tilde \Phi$ has  not the meaning
of {\it moduli } space), the characteristic excitation energies associated with $\tilde \Phi$ are of the same order as the
energies associated with $C_j$, and to ignore $\tilde \Phi$ would be inconsistent.

For positive $k$, the index of the Hamiltonian (\ref{HeffN2}) is given, again, by the 2-dimensional flux of magnetic field
as in (\ref{indCecotti}). But, in contrast to what happens in ${\cal N}=1$ theory, it does not change sign for negative $k$. We derive
for $SU(2)$
 \be
\label{IN2preWeyl}
I_{\rm pre-Weyl}^{{\cal N} = 2} \ =\ 2|k| \ .
 \ee
When imposing the condition of Weyl-invariance we are left with only $|k| + 1$ bosonic vacuum states.

Loop corrections do not change this result because, for ${\cal N} =2$ theory, fermion and gluon loop contributions
in the renormalization of $k$ and in the magnetic field flux cancel out. A generalization to higher $N$ is straingforward.
The final result for the index coincides with (\ref{ItreeSUN}), but without the factor $(-1)^{N-1}$. When $k \neq 0$, supersymmetry
is unbroken. 

If $k=0$, the pure ${\cal N} = 2$ $3d$ SYM field theory is known to involve the ``runaway'' vacuum: the degeneracy of the vacuum
valley is lifted by a superpotential generated by instantons such that the minimum of energy is achieved at infinitely large field values
\cite{runaway}. It would be interesting to understand how this is reflected in the finite-volume version of the theory.

\vspace{2mm}

\centerline{${\cal N} =3$}

\vspace{2mm}

The Lagrangian of ${\cal N}=3$ theory, involves four fermions,  $\psi_{f=1,2,3}$ and $\chi$. The fermion $\chi$ has the mass of opposite sign
compared to that of $\psi_f$. Besides, there are three real adjount scalars $\Phi_f$. The effective Hamiltonian (for $SU(2)$ theory) 
has the form
   \be
    \label{HeffN3}
   \frac {L^2}{g^2} H \ =\ \frac 1{2} \left(P_j - \frac {B}2 \epsilon_{jk} C_k \right)^2 + 
\frac {\tilde P^2_f
+  B^2 \tilde \Phi^2_f}2 + 
 \frac {B}2  (\psi_f \bar\psi_f - \bar\psi_f \psi_f)
- \frac B2 (\chi \bar\chi - \bar\chi \chi)
    \ee
The pre-Weyl index of this Hamiltonian is 
   \be
\label{IN3preWeyl}
I_{\rm pre-Weyl}^{{\cal N} = 3} \ =\ -2|k| 
 \ee    
(the vacuum states involve the factor $\chi$ and have negative fermion charge).
After imposing the Weyl invariance condition, only $|k| + 1$ fermion vacuum states are left. 
Loop corrections do not modify this value, because
positive (at positive $k$) shift of the magnetic flux due to gluon and $\chi$  loops  cancels its negative shift due to
$\psi_f$ loops. The final result for the index with arbitrary $N,k$  coincides 
up to a sign
\footnote{An overall sign of the index is, of course, a convention.}
 with the result in ${\cal N} =2$ theory.

\section{Discussion}
\setcounter{equation}0

\vspace{2mm}

\centerline{\it A paradox}

\vspace{2mm}

The problem of calculating the index (i.e. the number of vacuum states) in SYMCS theory 
is closely related to the problem of calculating
the total number of states in the topological pure CS theory. Indeed, the canonical momenta derived from the Lagrangian 
\be
\label{pureCS}
{\cal L}_{CS} \ =\ -\kappa \epsilon_{jk} \left[ {\rm Tr} \{A_j \dot A_k\} + {\rm Tr} \{ A_0 F_{jk}\} \right] 
 \ee 
are
\be
\label{canmom}
\Pi_j^a \ =\ \frac \kappa  2 \epsilon_{jk} A_k^a \ .
 \ee 
There is no time derivatives in the RHS,  and we obtain thereby a set of second class (they do not all commute) constraints
$G_j^a = \Pi_j^a - (\kappa/2) \epsilon_{jk} A_k^a = 0$ supplemented by the gauge constraints $F_{jk}^a = 0$. When quantizing, 
we have to replace, as  usual $\Pi_j^a \to -i\delta/(\delta A_j^a)$ and impose  the conditions 
 \be 
\label{vtorojrod} 
(\hat G_1^a + i \hat G_2^a) \Psi[A] \ = \ 0  \ \ \ \ \ \ \ {\rm or} \ \ \ \ \ \ \ (\hat G_1^a - i \hat G_2^a) \Psi[A] \ =\ 0
 \ee
on the wave functions (one   has to use a kind of  Gupta-Bleiler quantization procedure here and implement only a {\it half} of $G_j^a$ 
 \cite{indus}).

On the other hand, it is not difficult
to see that the supercharges of the SYMCS model (\ref{LN1})  can be represented as  
 \be
\label{supercharges}
   Q &=& \frac {g^2}2 \int d^2x \, \lambda^a_- \, G^a_{1-i2}\ , \nonumber \\
   \bar Q &=& \frac {g^2}2   \int d^2x \, \lambda^a_+ \, G^a_{1+i2} 
 \ee
($ \lambda_\pm \equiv \lambda_1 \pm i \lambda_2$). For positive $k$, ground states are bosonic and are annihilated
by $\bar Q$ in a trivial way. The condition $Q|\Psi\rangle =0$ is equivalent to the set of constraints 
$\hat G^a_{1-i2} \Psi = 0$. For negative $k$, the condition  $\bar Q|\Psi\rangle =0$ is equivalent to the set of constraints
$\hat G^a_{1+i2} \Psi = 0$. 

It is not surprising therefore that our results
[like (\ref{nashindextree})] for the tree-level index coincide with those derived earlier for 
pure CS theories. 
A conventional way to count the number of states in CS theories is to use their relationship \cite{WitCMP} to 2d WZNW
theories \cite{WZNW}, the correspondance between WZNW theories and conformal theories, 
and the full conformal machinery \cite{Shifconf}. But it can also be done by  resolving directly the constraints 
(\ref{vtorojrod}) \cite{Eli,Hotes}. 

The SYMCS theory in question involves, however, also the supersymmetric YM part in the action, which might affect the index.
Witten suggested that only the fermion part of this action does. His logic was the following \cite{private}.
Let us integrate over the fermions (after which the effective coupling is shifted according to $k \to k - c_V/2$) 
and obtain a purely bosonic theory. At low energies, this is the pure CS theory. 
It involves also the YM term and still higher derivative terms. Though these terms are irrelevant at low energies in a sense that the dynamics
depends exclusively on the  lowest dimension CS term in the Wilsonian effective Lagrangian, they can affect
the coefficient of this term. However, in contrast to what happens, e.g., in a conventional $4d$ YM theory 
supplemented by  a higher-derivative term $\sim {\rm Tr} \{ F D^2 F\}/M^2$, where the effective low-energy YM coupling constant involves
a logarithmic dependence of $M$, in this case, renormalization of $\kappa$ does not depend on the coefficient $1/g^2$ of the YM term.
Moreover, it does not depend on the {\it form} of the higher derivative terms, the result 
\be
\label{bosshift}
k_{\rm bosonic}\  = \ k_{\rm tree} + c_V
 \ee 
bein robust with respect to these details. One can therefore consider this shift as an immanent feature of pure CS 
theory, with quantum effects taken into account. Indeed, the shift $k \to k+c_V$ appears in many exact formulae, like those for the 
energy-momentum tensor
or Wilson loops expectation values, etc \cite{WitCMP,Shifconf}. On the other hand, this shift does not show up in the formula
 (\ref{nashindextree}) for the number of states in CS theory (on the conformal side, it is the number of so called {\it conformal blocks}).
 Thus, concludes Witten, one should not take into account the renormalization of $k$ due to bosonic loops. The known pattern of the
exact solution of pure CS theory displays that bosonic loops are present, indeed,  they affect $\kappa$ and other quantities, 
but do not affect the number of states.     
 
This reasoning looks OK. Besides, supersymmetry breaking at small $k$ that it implies follows also from heuristically suggestive
$D$-brane constructions \cite{Ohta}.
However,  it is somewhat formal, relying heavily on the correspondence with conformal theories and 
exact results there. It does not give a clear physical picture of what really happens. Our method, consisting in explicit evaluation
of the low-energy Hamiltonian in finite volume, gives such physical picture, but, surprisingly, the result of this analysis is different
- bosonic loops {\it do} contribute to the index. This is an obvious paradox, which should be resolved somehow. Being unable now to make
essential comments  
on the conformal way of reasoning, let us try to see whether one can modify {\it our} prediction following from the  analysis of 
the effective finite-volume Hamiltonian.
\footnote{This analysis is done in the region $g^2L \ll 1$, while the pure CS dynamics refers to the limit
$g^2L \gg 1$. The standard wisdom tells, however, that the number of vacuum states should not depend on $L$.} 

 One of the places in our proof, which {\it might} involve a loophole
\footnote {It goes without saying that there are  {\it a lot} of loopholes if mathematical standards of rigour are applied. 
This concerns this paper, the paper \cite{Wit99}, and, basically, all other papers on quantum field theory published in  physics journals.}
is the following.
In Appendix, we evaluated accurately the contribution of the fermion loop into the effective finite-volume BO Hamiltonian and
confirmed that, as far as the expression for the induced magnetic field is concerned, the simple rules (\ref{pnC}) 
work and, as a result, the flux of induced magnetic field is rigidly connected with renormalization of $\kappa$ in infinite volume.
This generalizes to bosonic loops if the latter are evaluated with the simplistic infrared regularization (\ref{massterm}).
 It is difficult to imagine that the results may depend here on the regularization, but we cannot logically exclude now that, when
accurate calculations are done in full SYMCS theory, and the extra terms in the action involve a couple of derivatives, 
the recipe (\ref{pnC}) breaks down for bosonic loops. As a result, the flux of the induced magnetic field might be zero in spite of
 the nonvanishing renomalization of $\kappa$ in the infinite-volume theory... 
To patch the hole, such an accurate calculation should be performed.

Another potential source of trouble is the fact that the BO approximation we use breaks down near some special points 
(fixed points of Weyl transformation) on the flat connection space [see the footnote \ref{fixed}]. 
This allows one to suspect the presence of some extra contributions in the index that we did not take into 
account. They might be {\it (i)} extra one-loop contributions and/or {\it (ii)} higher-loop contributions. Speaking of the latter, 
we have not rigourously excluded their presence for higher-rank groups [see the footnote \ref{nohigherloop}], but,
for $SU(2)$, we did. (We remind: for $SU(2)$, higher-loop contributions (if any) should involve inverse powers of $k$, this is not
allowed  at large $k$ and hence the coefficient should be zero for any $k$). Speaking of the former, we analyzed accurately 
only a possible correction to the index due to renormalization of the metric and showed that it vanishes. However, there are many other corrections
with four and more derivatives in the Lagrangian. As the index is a topological quantity, it is difficult to imagine that something else besides
the (generalized) flux might contribute, but, again, it is not a mathematical theorem. There is a logical possibility that something 
queer, like higher-derivative terms, contributes to the index and this cancels the contribution of the  flux induced by the bosonic loop. 

 The third possibility is the following. The Weyl group $W$ has a natural $Z_2$ gradation involving even and odd elements. 
(For example, for $SU(N)$, the Weyl
group  $S_N$ involves even and odd permutations.)  
Imagine now that, for some reason, we should have picked up not Weyl invariant wave functions, but rather {\it Weyl antiinvariant} ones, i.e.
the functions that are invariant under the action of even elements of $W$ and change sign under the action of odd elements
\cite{Eli,Axelrod}. Weyl antiinvariant wave functions can be represented as [cf. Eq.(\ref{sumWeyl}]
  \be
\label{sumantiWeyl}
\Psi \ =\ \sum_{\hat{x} \in W} \hat{x} \Psi_{\vecb{w}_n} \, P(x) \ ,
 \ee    
where $P(x) = \pm 1$ depending on whether the element $x$ is even or odd. 

It is not difficult to see then that the number of such Weyl antiinvariant states is equal to the number
of points in the Weyl alcove {\it excluding} the points on its boundary. Indeed, the latter are invariant with respect to a 
$Z_2$ subgroup of the Weyl group that involves the unity and some odd element of $W$ of second order. 
[ For the Weyl alcove of $SU(3)$ depicted in Fig. 2, this odd
element is one of the permutations (12), (13), or (23) --- see the footnote \ref{granicy}]. A glance at Fig.2 tells that, for $k=4$, there are
 only three points in the interior of the alcove. And this coincides with the number of points in the Weyl alcove for $k=1$ counted in the conventional
way (with inclusion of the boundary). One can note now that $1 = 4-3$, i.e., for $k=4$,
 {\it the number of Weyl antiinvariant states for the $SU(3)$ effective Hamiltonian
that takes into account the contribution of the gluon loops bringing about the shift $k \to k+3$ is equal to the number of Weyl invariant states for the
unshifted Hamiltonian}. A pure geometric inspection of larger triangles and multidimensional tetrahedrons displays that this pattern also holds
for all $k$ and $N$. For higher unitary groups, Weyl antiinvariance condition ``unwinds'' the gluon loop shift $k \to k+N$. We enjoyed observing
this
 also for symplectic groups $Sp(2r)$ (where counting the points in the interior of the alcove unwinds the shift $k \to k+c_V = k + r + 1$) 
and for $G_2$.
Indeed, looking at the Weyl alcove for $G_2$ in Fig. 2, one observes that, for $k =4$, only one state is left, and this corresponds to unwinding
$k \to k-4 = k - c_V[G_2]$. This theorem can be proven for an arbitrary group  \cite{Eli,Axelrod}: Weyl antiinvariance requirement amounts 
always to the negative shift 
$k \to k - c_V$ that compensates the shift (\ref{bosshift}) due to gluon loops. 
In other words, by imposing Weyl antiinvariance requirement on the wave functions, 
we would reproduce Witten's result. The problem is, however, that we do not see a reason to do that in the framework of our approach.

\vspace{2mm}

\centerline{\it Going down onto the quotient.}

\vspace{2mm}

We calculated the index by studying the dynamics on the moduli space of all (not necessarily gauge equivalent) 
Abelian flat connections and imposing then the Weyl invariance condition on the quantum states. The advantage of this 
approach is simplicity of such moduli space - just the product $T \times T$ of two copies of the  maximal torus of the gauge group. 
An alternative approach is to  factorize $[T \times T]$ over the Weyl group
$W$ at the classical level and study the dynamics on the (more complicated) moduli space thus obtained. This is the way the index
 was calculated
in Sect. 3 of Ref.\cite{Wit99}. This calculation uses a bunch of nontrivial mathematical facts, which we have understood (with the help
of colleages mathematicians) only partially.
 Still, we have decided to make here few explanatory comments, which might be useful for an unsophisticated physicist reader
 who shares with the author his mathematical illiteracy.

\begin{figure}[t]
\begin{center}
\includegraphics[width=2in]{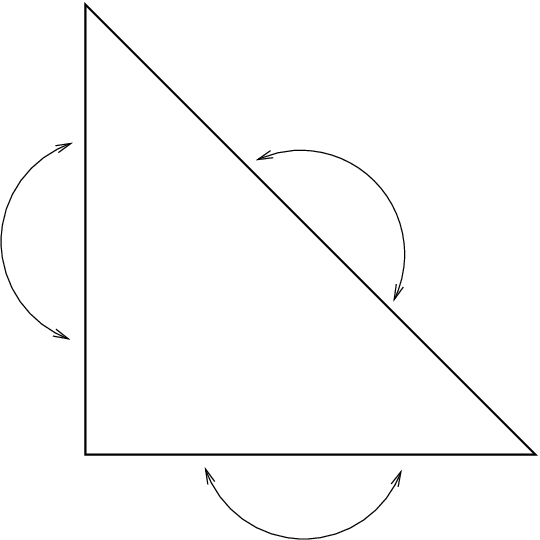}
\end{center}
\caption{$T^2/Z_2 = S^2$.}
\label{konvert}
\end{figure}

The first nontrivial fact is  
  that the moduli space ${\cal M} = [T^{\rm max} \times T^{\rm max}]/W$ of gauge equivalent classes of flat $SU(N)$ connections
on $T^2$ is $ {\mathbb C \mathbb P}^{N-1}$ \cite{FMW}. 
In fact, the proof of a similar statement for {\it symplectic} groups (that  ${\cal M}_{Sp(2r)} = [T^{\rm max}_{Sp(2r)} 
\times T^{\rm max}_{Sp(2r)}]/W_{Sp(2r)} = {\mathbb C \mathbb P}^r $) is much simpler.  
Consider first $Sp(2) = SU(2)$. In this case, $T^{\rm max}$ is just a circle and the Weyl group is
$Z_2$. Then ${\cal M}$ is a set of points $(x,y)$ identified by periodicity $(x,y) \equiv
(x+1,y) \equiv (x,y+1)$ and by simultaneous Weyl reflection $(x,y) \equiv (-x,-y)$. This gives a triangle
with glued edges
depicted in Fig.\ref{konvert}: the points symmetric with respect
to the middles  of the edges are identified. An ``envelope'' thus obtained is topologically equivalent
to $S^2$. 

For $Sp(2r)$, $T^{\rm max}$ is a direct product of $r$ such circles. The Weyl group has $2^r\cdot r!$ 
elements including reflections on each such circle and their permutations. It is clear then that
  \be
\label{rsfer}
{\cal M}_{{\rm Sp}(2r)} \ =\ \frac {\frac {S^1 \times S^1}{Z_2} \times \cdots  \times \frac {S^1 \times S^1}{Z_2}}{S_r} 
\ =\ \frac {S^2 \times \cdots \times S^2}{S_r}\ 
 \ee
with $r$ \ $S^2$ factors. Introduce a complex structure on each  factor. A point in ${\cal M}$ can be represented as an 
unordered set of $r$ complex numbers $(z_1, \ldots, z_r)$. One can represent this set as a set of roots of some
polynomial of order $r$ and map the set of all such sets  
to the set of all complex polynomials of degree $r$ factorized over multiplication by a complex factor $\lambda$. 
Bearing in mind that a polynomial of degree $r$ is represented by a set of $r+1$ its coefficients, we derive
${\cal M} \equiv  \mathbb C  \mathbb P^r$, as promised. 

Let us go over to unitary groups. The maximal torus of $SU(N)$ is a set of matrices diag$(e^{i\alpha_1},\ldots, e^{i\alpha_N})$
with $\sum_{l=1}^N \alpha_l = 0$.
The product of two such tori can be represented as the space of sets $\{z_1,\ldots,z_N\}$, where $z_l = \alpha_l + i\beta_l$ belongs
to $T^2$ [$\alpha_l, \beta_l \in (0,2\pi)$] and $\sum_l z_l = 0$. The Weyl group permutes $z_l$. Thus, ${\cal M}_{SU(N)}$ is 
a set of unordered $N$-tuples on $T^2$ that add to zero. Similarly to what was done in the case of $Sp(2r)$, such $N$-tuple
can be  represented by  meromorphic elliptic functions defined on $T^2$ that have simple zeroes at $N$ selected
points and the pole of $N$-th order at zero.
\footnote{If one of the coordinates $z_l$ is zero, the pole is of order $N-1$, if two of them are zero, the pole is
of order $N-2$, etc.}
 There is a one-to-one correspondence between these $N$-tuples and 
the classes of such functions $F(z)$ with identification $F(z) \equiv \lambda F(z)$. It is a known mathematical fact that
the space of all such elliptic functions is a vector space of complex dimension $N$. Bearing in mind the identification
with respect to multiplication by $\lambda$, the projective space $\mathbb C \mathbb P^{N-1}$ arises.

Witten then relates the index to a certain topological invariant of $\mathbb C \mathbb P^{N-1}$ associated with the presence 
of extra Abelian gauge field on this manifold. We do not want to go  into further details (bearing especially in mind that we 
do not understand this question completely), but we would like to mention here that an {\it elementary} calculation of this
invariant 
\footnote {It uses the explicit form of the metric on $\mathbb C \mathbb P^{N-1}$ and directly counts the normalized states 
annihilated by the properly chosen second class constraints [cf. Eq.(\ref{vtorojrod})].} was performed in \cite{IMT}. The number 
of the states depends at the tree level on the parameter $k$ and is given by (\ref{nashindextree}). As was discussed above, 
Witten suggests that $k$ should be shifted due to fermion loops to $k - c_V/2$, while our analysis suggests the positive 
shift $k \to k + c_V/2$. 

\vspace{2mm}

\centerline{\it Strings and walls}

\vspace{2mm}

 The last rather confusive issue that we want to discuss here are the arguments of Ref.\cite{Vafa} relating the Witten index 
in 3d SYMCS theory at the level
$k$ to the multiplicity
of {\it domain walls} in ${\cal N} = 1$ $4d$ SYM theory with $SU(k)$ gauge group. 
The standard reasoning displaying the appearance of these walls is the following.
The tree Lagrangian of this theory involves axial $U(1)$ symmetry. Like in QCD, this symmetry is anomalous, being broken by instantons. An 
instanton    
possesses   $2k$ gluino zero modes, the 't Hooft determinant involves the factor $\sim \lambda^{2k}$, and that means that the discrete
$Z_{2k}$ subgroup of the axial $U(1)$ group remains unbroken. This discrete symmetry is further {\it spontaneously} broken down to $Z_2$, with
the phase of the gluino condensate,
 \be
\label{gluinocond}
 \langle  \lambda \lambda \rangle_l \ =\ \Sigma e^{2\pi i l/k}\ , \ \ \ \ \ \ \ \ \ l = 1,\ldots,k \ , 
 \ee
playing the role of the order parameter of this breaking \cite{gluino}. This implies the existence of $k$ distinct vacua and
domain walls separating them \cite{Dvali}. There are domain walls of different kind interpolating between the vacua with phase differences 
$p = l-l' = 1,\ldots, k-1$. For given $k,p$, there are several different domain walls, their multiplicity being evaluated (by brane methods) 
in \cite{Vafa} as 
 \be  
\label{wallmult}
 \#_{\rm walls}(k,p) \ =\ \left( \begin{array}c k \\ p \end{array} \right) \ .
 \ee
Based on certain $D$-brane and duality arguments,  Acharya and Vafa relate this number to the number of vacuum 
states in ${\cal N} =2$ $3d$ SYMCS $SU(p)$ theory at level $k$
(the main idea is that the effective theory { on} the domain wall {\it is} in fact a $3d$ SYMCS theory). And this relation
holds if using the ${\cal N} =2$ generalization \cite{Ohta}
 \be
\label{N2poWit}
 I \ =\ \left( \begin{array}c k-1 \\ p-1 \end{array} \right)
 \ee
 of  original Witten's formula (\ref{IndpoWittenu}) and {\it not} our formula (\ref{nashindex}) !       
\footnote{The exact agreement between (\ref{wallmult}) and (\ref{N2poWit}) is achieved, if taking into account 
the presence of an extra $U(1)$ factor in the effective theory. As a result, the number of walls is given by the
$SU(p)$ index (\ref{N2poWit}) multipled by the factor $k/p$ \cite{Vafa}, which coincides with (\ref{wallmult}).}

Even though this agreement looks to be rather remarkable, it is not conclusive enough in the framework of our restricted 
rules
of the game, where only pure field theory reasoning is admissible, and duality arguments are not.

SYM theory is a theory with strong
coupling, and it is difficult to perform a honest study of domain walls there and count their number. 
The only ``braneless'' way to do it  is to modify
the theory by adding there extra fundamental matter multiplets \cite{KSS}. 
If the matter fields are light enough, one can integrate over all other
degrees of freedom to obtain the effective ADS Lagrangian \cite{ADS}. It is a Lagrangian of Wess-Zumino type, with superpotential
 involving a special instanton-generated term. It has, indeed, $k$  different vacua, and the classical solutions describing
different domain walls can be explicitly constructed and counted, their number being given by Eq.(\ref{wallmult}). But it is not 
evident that the number of walls in the pure SYM theory should be the same. 
The latter can be achieved from the weakly coupled theory with light
matter fields by increasing their mass. If the mass becomes very large, these fields decouple. 
{\it If} the number of walls is not changed under
  such deformation, the counting (\ref{wallmult}) works also for pure SYM. 

This condition seems not to be fulfilled, however. In Refs.\cite{stenki,tenac}, this very question was studied in the framework of the
Taylor-Veneziano-Yankielowicz Lagrangian \cite{TVY} involving on top of  matter superfields also the chiral superfield $S$, which takes 
effectively into account the gluon and gluino degrees of freedom. 
\footnote{The TVY Lagrangian has correct symmetry properties, but it is not a Wilsonean effective Lagrangian, and one cannot be sure that the results
obtained in the  TVY framework hold also for the full SYM theory. Anyway, it is the {\it only} known to us field theory 
method to study domain walls in strongly coupled regime.}

This study has revealed that, when the mass is increased, a chain of phase transitions 
(or rather bifurcation points) occur such that  most of the walls
disappear at large masses. 
\footnote{Disappearance of walls in the pure SYM theory might be associated the fact that the standard interpretation with spontaneous breaking
of discrete chiral symmetry is actually questionable. The pure SYM theory, unlike a theory with fundamental matter, admits not only 
instanton 
Euclidean configurations with integer topological charge, but also configurations with fractional charge. Such configurations 
('t Hooft torons \cite{tHooft}) 
are certainly there
in a theory defined on a spatial torus with twisted boundary conditions [cf. a remark after Eq.(\ref{quantkap})]. And then the 
phase of the fermion condensate $\langle \lambda \lambda \rangle$ in Eq.(\ref{gluinocond}) is not an order parameter, but plays 
the same role as the vacuum angle $\theta$ - it should be chosen once for ever, and there are no physical walls connecting  vacua
with  different $\theta$.
See  Refs.\cite{KSS,tenac} for discussion of this controversial issue.}     
For example, for $k=2$, both walls disappear. For $k=3$ only two ``tenacious'' walls out of  three are left, etc.
In other words, the counting (\ref{wallmult}) works for the ADS Lagrangian, but probably does not work for SYM theory. 
Bearing this in mind, the agreement between Eq.(\ref{wallmult}), which does not count correctly the number of walls, and 
Eq.(\ref{N2poWit}), which
is not a correct value of the $3d$ index, looks really misterious... 

We are indebted to E. Witten for profound illuminating discussions and many valuable remarks. 
We aknowledge also useful discussions with  B. Feigin, A. Gorsky, E. Ivanov, 
A. Pajitnov, V. Rubtsov, and S. Theisen.

\section*{Appendix. Magnetic flux induced by loops.}
\setcounter{equation}0
\def\theequation{A.\arabic{equation}}

\vspace{1mm}

\centerline{$SU(2)$}

\vspace{2mm}

The formula (\ref{DBferm}) for the induced magnetic field on the dual torus is very natural and follows
{\it almost} directly from (\ref{PioF}) and the rules (\ref{pnC}). However,
 this simple correspondence is formulated for the magnetic field ${\cal B}$, while 
the bosonic part of the effective Lagrangian involves vector-potential ${\cal A}$ rather than ${\cal B}$, and
the formula for ${\cal L}_{\rm eff}$ is more complicated. Because of this and because of the controversy concerning the bosonic
loop contribution,  we decided to make here some explanatory comments. 

Consider the fermion contribution. To find the correction to the effective Lagrangian, we have to evaluate
the fermion loop in finite volume in external  background field
 \be
\label{Afon}
\vecb{C}(\tau) \ =\ \vecb{C}_0 + \vecb{E}\tau  \ , 
 \ee
where $\tau$ is Euclidean time
(to evaluate the graphs, we are going to perform, as usual, Wick rotation, etc.). 
For any multileg graph  in the expansion of Tr $\ln  (i {\cal D}\!\!\!\!/ - m )$, we have 
thus to insert such $\vecb{C}(\tau)$ in each leg and keep
 only  {\it linear} in $\vecb{E}$  terms.
\footnote{We are hunting for the structure $\sim \dot{\vecb{C}} {\vecb{\cal A}}(\vecb{C})$ in ${\cal L}^{\rm eff}$. 
The quadratic in $\dot {\vecb{C}}$ 
terms give corrections to the metric, there are also cubic and all higher-order terms, but they do not affect the index, 
as was discussed in Sect. 3.} 
The way the calculations are done here \cite{Becker} is very much parallet to the technique of calculations in background nonperturbative
Euclidean $4d$ fields developped in \cite{tehnika} and based on the gauge choice\cite{fixed}  $(x-x_0)_\mu A_\mu = 0$ leading to 
  \be
\label{fix}
 A_\mu(x)  \ = \ \int_0^1 s ds \, (x-x_0)_\nu  F_{\nu\mu}[x_0 + s(x-x_0)] = \nonumber \\
 \frac 12(x-x_0)_\nu  F_{\nu\mu}  + \frac 13 (x-x_0)_\nu (x-x_0)_\alpha  {\cal D}_\alpha  F_{\nu\mu} + \ldots
 \ee
This gauge is not translationally invariant, but the physical results must not (and do not) depend on the choice of the ``fixed
point'' $x_0$. This choice is in our hands. Likewise, the point $\tau_0$ at which the linear term in the decomposition (\ref{Afon})
vanishes, is a convention. We will choose $\tau_0 = 0$  coinciding with the position of one of the legs in the graphs.

\begin{figure}[t]
\begin{center}
\includegraphics[width=2.5in]{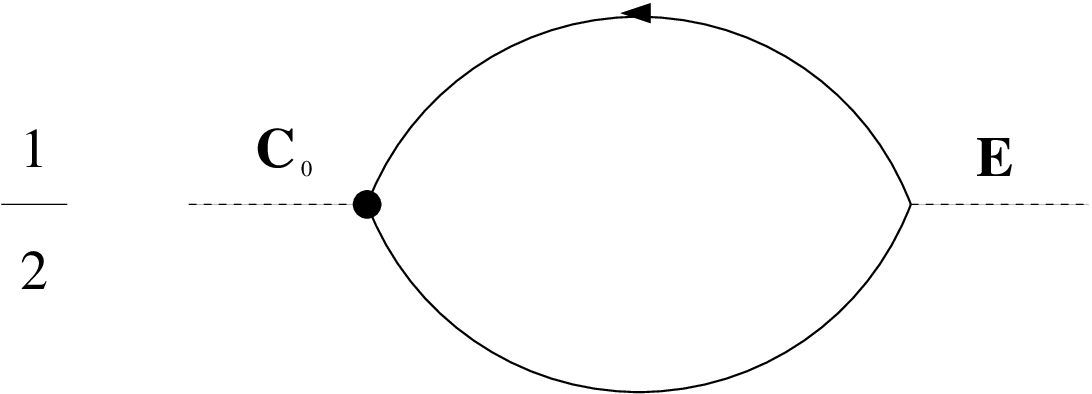}
\end{center}
\caption{$EC$ contribution to ${\cal L}^{\rm eff}$.}
\label{2legEA}
\end{figure}

\begin{figure}[t]
\begin{center}
\includegraphics[width=5in]{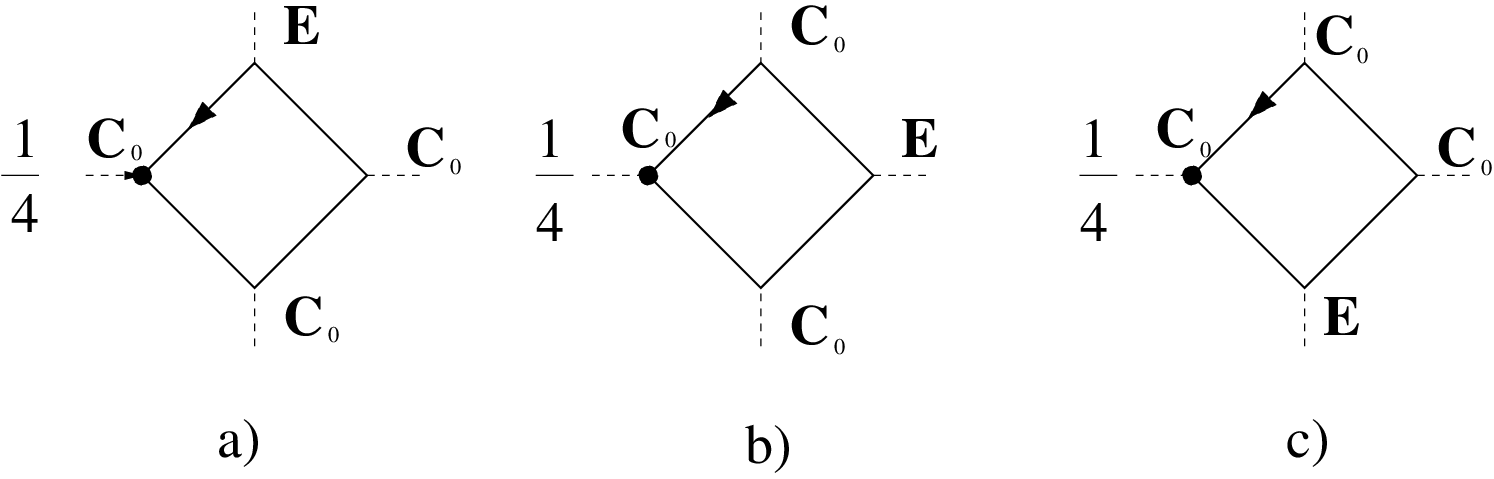}
\end{center}
\caption{$ECCC$ contribution to ${\cal L}^{\rm eff}$.}
\label{4legEAAA}
\end{figure}

The graphs with an odd number of legs vanish, and we have to consider only the graphs with even number of legs.
There is only one two-leg graph depicted in Fig.\ref{2legEA}. The factor 1/2 coming from the expansion of 
$\ln (i {\cal D}\!\!\!\!/ - m )$ is displayed. The blob marks the ``fixed point'' - the vertex at $\tau = 0$. At this point, 
one can plug
only the constant part of $\vecb{C}(\tau)$, the linear term vanishes there. There are three nonzero 4-leg graphs depicted in 
Fig.\ref{4legEAAA}. The expansion factors $1/4$ are displayed.

The $EC$ contribution to the effective Lagrangian is proportional to 
 \be
\label{EC}
{\cal L}^{\rm eff}_{\rm Fig. \ref{2legEA}} \ \propto \frac 1{2} \sum_{\vecb{n}} C_j E_k \int \frac {d\epsilon}{2\pi}
{\rm Tr} \left\{ \gamma_j G \gamma_k \frac {\partial G}{\partial \epsilon} \right\} \ ,
 \ee
where $G(\epsilon, 2\pi\vecb{n}/L)$ is fermion Green's function, and we took care to display explicitly
the factor 1/2 coming from the expansion of the logarithm, but not other numerical factors. We have also suppressed
from now on the subscript $_0$ for $\vecb{C}$.  
For the graphs in Fig.\ref{4legEAAA}, the factor $\tau$ multiplying $E_k$ goes over into the  operator $\partial/(\partial \epsilon)$ 
acting on all Green's functions between the point where $E_k\tau$ is inserted and the blob in, say, the clockwise direction
\cite{tehnika}. For example, the graph in Fig.\ref{4legEAAA}b gives
   \be
\label{ECCC}
{\cal L}^{\rm eff}_{\rm Fig.\ref{4legEAAA}b} \ \propto \frac 1{4} \sum_{\vecb{n}} C_j C_l E_k C_p
 \int \frac {d\epsilon}{2\pi}
{\rm Tr} \left\{ \gamma_j G \gamma_l G \gamma_k \frac {\partial}{\partial \epsilon} (G \gamma_p G) \right\} \ ,
 \ee
Again, only the expansion factor 1/4 is explicitly displayed. The 6-leg graphs $\sim EC^5$ involve the expansion
factor 1/6, etc. 

\begin{figure}[t]
\begin{center}
\includegraphics[width=2.5in]{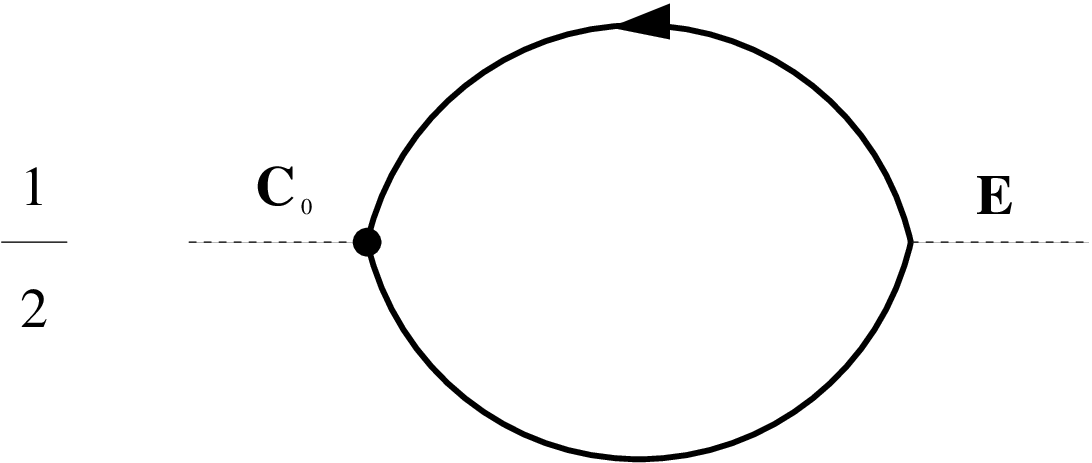}
\end{center}
\caption{Two-leg graph with thick Green's functions.}
\label{exact}
\end{figure}  

To resum all such contributions, let us compare the expressions (\ref{ECCC}) etc. to the corresponding terms in the
expansion of the graph in Fig. \ref{exact}, where the thick lines stand for Green's functions in the {\it constant} background 
$\vecb{C}$. These expansion terms have the same structure as in Eq.(\ref{ECCC}), but the coefficients 1/4, 1/6, etc are replaced
by a universal combinatorial prefactor 1/2. To find ${\cal L}^{\rm eff}$ to any order in $\vecb{C}$, we have thus to
take the expression
 \be
m  \epsilon_{jk} C_j E_k  \sum_{\vecb{n}}\, \int_{-\infty}^\infty \frac {d\epsilon}{2\pi} \,
\frac 1 {\left[ \epsilon^2 + \left( \frac {2\pi \vecb{n}}L - \vecb{C} \right)^2 + m^2 \right]^2 }
\ee
for the graph in Fig.\ref{exact}, expand it in $\vecb{C}$, and divide the $2p$-th term of this expansion by  $p+1$. 
We obtain
 \be
\label{LECn}
    {\cal L}^{\rm eff}_{EC\cdots C} \ =\ -E_k {\cal A}_k(\vecb{C})
\ee
with
 \be
\label{AviaB}
{\cal A}_k(\vecb{C}) \ =\  - 2m \epsilon_{jk} C_j 
 \sum_{\vecb{n}} \int_0^1 s ds \int_{-\infty}^\infty \frac {d\epsilon}{2\pi} 
\, \frac 1 {\left[ \epsilon^2 + \left( \frac {2\pi \vecb{n}}L - 
s \vecb{C} \right)^2 + m^2 \right]^2}\ .
 \ee
This is nothing but Fock-Schwinger gauge representation (\ref{fix}) for the vector potential via magnetic field.
We thus arrive at the result (\ref{DBferm}) for $\Delta {\cal B}^F(\vecb{C})$.  

An {\it explicit} evaluation of   $\Delta {\cal B}^B(\vecb{C})$ in SYMCS theory is technically  
more involved. In the background field method, there are two types of vertices with single and double external field insertions.
In addition, the expression for the gluon propagator is more complicated. What we can easily do, however, is to calculate the induced
magnetic field in the model where the YM term in the action is replaced by the gluon mass term (\ref{massterm}). Then the 
action is exactly  the same  as for the fermions and the results are also exactly the same up to the factor -2,
 \be
\label{B-2F}
 \Delta {\cal B}^B (\vecb{C}) = -2\Delta {\cal B}^F (\vecb{C})\ .
 \ee
 In view of the controversy discussed in the paper (whether gluon loops are relevant or not), 
it would make sense to perform this calculation 
with the ``honest'' YM action.  It is difficult to imagine, however, that some other
result than (\ref{B-2F}) would be obtained. 
At $\vecb{C} = 0$, the equality (\ref{B-2F}) is manifest with any regularization.   

\vspace{1mm}

\centerline{$SU(3)$}

\vspace{2mm}

For an arbitrary group, the loop-corrected expression for the generalized flux (\ref{indCecotti3}) is obtained
from (\ref{ISUN}) by substituting $ k \to k + c_V/2$. Let us show  how this comes about
 in the simplest nontrivial 
$SU(3)$ case. Consider the fermion contribution. The charged [with respect to the background 
$(\vecb{A}^3, \vecb{A}^8)$] fermions circulating in the loops, like
in Figs.\ref{2legEA},\ref{4legEAAA}, can be directed along three root vectors of $SU(3)$, 
$\psi^{1\pm i2}, \psi^{4\pm i5}$, and $\psi^{6\pm i7}$. They give contributions depending on the roots
$\vecb{A}^3, \ (-\vecb{A}^3 + \sqrt{3} \vecb{A}^8)/2$ and $-(\vecb{A}^3 \sqrt{3} + \vecb{A}^8)/2$, correspondingly.  

The total 1-loop contribution to the effective Lagrangian is expressed as
    \be
\label{LeffSU3}
    {\cal L}^{\rm eff} \ =\ -E^3_k {\cal A}^3_k - E^8_k {\cal A}^8_k
   \ee  
with 
\be
\label{calA38}
{\cal A}^3_k &=& {\cal A}_k(\vecb{A}^3) - \frac 12 {\cal A}_k\left (\frac{-\vecb{A}^3 + \sqrt{3} \vecb{A}^8}2 \right) 
- \frac 12 {\cal A}_k\left( -\frac{\vecb{A}^3 + \sqrt{3} \vecb{A}^8}2 \right) \nonumber \\
{\cal A}^8_k &=& \frac {\sqrt{3}}2 \left[ {\cal A}_k \left(\frac{-\vecb{A}^3 + \sqrt{3} \vecb{A}^8}2 \right) 
- {\cal A}_k\left( -\frac{\vecb{A}^3 + \sqrt{3} \vecb{A}^8}2 \right) \right]\ ,
 \ee
  where a universal function ${\cal A}_k(\vecb{C})$ is taken from (\ref{AviaB}). Let us add now the contribution from gluon loops
(this amounts to changing sign of ${\cal A}_k$) and calculate ${\cal B}^{ab}$ and its determinant.
We obtain
 \be
\label{detab}
\det\|{\cal B}^{ab} \| \ =\  (\kappa L^2)^2 + \kappa L^2 \left[{\cal B}(\vecb{A}^3) + 
{\cal B}\left(\frac{-\vecb{A}^3 + \sqrt{3} \vecb{A}^8}2 \right) + {\cal B} \left(-\frac{\vecb{A}^3 + \sqrt{3} \vecb{A}^8}2 \right)
\right] \nonumber \\
 + \frac 34 \left[ {\cal B}(\vecb{A}^3) {\cal B}\!\left(\frac{-\vecb{A}^3 + \sqrt{3} \vecb{A}^8}2 \right) + 
{\cal B}(\vecb{A}^3) {\cal B}\!\left(-\frac{\vecb{A}^3 + \sqrt{3} \vecb{A}^8}2 \right) + \right. \nonumber \\ 
 \left. {\cal B}\!\left(\frac{-\vecb{A}^3 + \sqrt{3} \vecb{A}^8}2 \right)  {\cal B}\!\left(-\frac{\vecb{A}^3 + \sqrt{3} \vecb{A}^8}2 \right)
\right]
 \ee
with 
\be
\label{Buniv}
 {\cal B}(\vecb{C}) \ =\ 2m  \int_{-\infty}^\infty \frac {d\epsilon}{2\pi} \sum_{\vecb{n}}
   \frac 1{\left[\epsilon^2 + \left( \frac {2\pi \vecb{n}}L - \vecb{C} \right)^2 +
   m^2 \right]^2} \ . 
\ee
Integrating (\ref{detab}) over $d\vecb{A}^3 d\vecb{A}^8$ within $T_{\rm max}[SU(3)] \times T_{\rm max}[SU(3)]$ gives $3(k+3/2)^2$.

\end{document}